\documentclass[]{aastex631}

\usepackage{float}
\graphicspath{{Figures/}}

\begin{document}

\title{The MUSCLES Extension for Atmospheric Transmission Spectroscopy: UV and X-ray Host-star Observations for JWST ERS \& GTO Targets}

\author[0000-0003-1369-8551]{Patrick R. Behr}
\affiliation{Department of Astrophysical \& Planetary Sciences, University of Colorado Boulder, Boulder, CO, 80309, USA}
\affiliation{Laboratory for Atmospheric and Space Physics, University of Colorado Boulder, Boulder, CO, 80303, USA}

\author[0000-0002-1002-3674]{Kevin France}
\affiliation{Department of Astrophysical \& Planetary Sciences, University of Colorado Boulder, Boulder, CO, 80309, USA}
\affiliation{Laboratory for Atmospheric and Space Physics, University of Colorado Boulder, Boulder, CO, 80303, USA}

\author[0000-0003-2631-3905]{Alexander Brown}
\affiliation{Center for Astrophysics and Space Astronomy, University of Colorado Boulder, Boulder, CO, 80309, USA}

\author[0000-0002-7119-2543]{Girish Duvvuri}
\affiliation{Department of Astrophysical \& Planetary Sciences, University of Colorado Boulder, Boulder, CO, 80309, USA}
\affiliation{Center for Astrophysics and Space Astronomy, University of Colorado Boulder, Boulder, CO, 80309, USA}

\author[0000-0003-4733-6532]{Jacob L.\ Bean}
\affiliation{Department of Astronomy \& Astrophysics, University of Chicago, Chicago, IL 60637, USA}

\author[0000-0002-3321-4924]{Zachory Berta-Thompson}
\affiliation{Department of Astrophysical \& Planetary Sciences, University of Colorado Boulder, Boulder, CO, 80309, USA}
\affiliation{Center for Astrophysics and Space Astronomy, University of Colorado Boulder, Boulder, CO, 80309, USA}

\author[0000-0001-8499-2892]{Cynthia Froning}
\affiliation{Department of Astronomy, University of Texas Austin, Austin, TX, 78712, USA}

\author[0000-0002-0747-8862]{Yamila Miguel}
\affiliation{Leiden Observatory/SRON, Leiden, Netherlands}

\author[0000-0002-4489-0135]{J. Sebastian Pineda}
\affiliation{Laboratory for Atmospheric and Space Physics, University of Colorado Boulder, Boulder, CO, 80303, USA}

\author[0000-0001-9667-9449]{David Wilson}
\affiliation{Department of Astrophysical \& Planetary Sciences, University of Colorado Boulder, Boulder, CO, 80309, USA}
\affiliation{Laboratory for Atmospheric and Space Physics, University of Colorado Boulder, Boulder, CO, 80303, USA}

\author[0000-0002-1176-3391]{Allison Youngblood}
\affiliation{NASA Goddard Spaceflight Center, Greenbelt, MD, 20771, USA}


\begin{abstract}
X-ray through infrared spectral energy distributions (SEDs) are essential for understanding a star's effect on exoplanet atmospheric composition and evolution. We present a catalog of panchromatic SEDs, hosted on the Barbara A. Mikulski Archive for Space Telescopes (MAST), for 11 exoplanet hosting stars which have guaranteed \textit{JWST} observation time as part of the ERS or GTO programs but have no previous UV characterization. The stars in this survey range from spectral type F4-M6 (0.14-1.57 M$_\sun$), rotation periods of ~4-132 days, and ages of  approximately 0.5-11.4 Gyr. The SEDs are composite spectra using data from the \textit{Chandra X-ray Observatory} and \textit{XMM-Newton}, the \textit{Hubble Space Telescope}, BT-Settl stellar atmosphere models, and scaled spectra of proxy stars of similar spectral type and activity. From our observations, we have measured a set of UV and X-ray fluxes as indicators of stellar activity level. We compare the chromospheric and coronal activity indicators of our exoplanet-hosting stars to the broader population of field stars and find that a majority of our targets have activity levels lower than the average population of cool stars in the solar neighborhood. This suggests that using SEDs of stars selected from exoplanet surveys to compute generic exoplanet atmosphere models may underestimate the typical host star's UV flux by an order of magnitude or more, and consequently, that the observed population of exoplanetary atmospheres receive lower high-energy flux levels than the typical planet in the solar neighborhood.
\end{abstract}



\section{Introduction}
\label{sec:intro}
The NASA Exoplanet Archive lists over 5000 confirmed exoplanets across nearly 4000 distinct systems. Estimates from \textit{Kepler} observations predict exoplanet occurance rates around F-, G-, and K-type stars of $\sim10-60\%$ \citep{kuminoto2020,fressin2013,traub2012,kopparapu2013} and up to 80\% or more for M dwarf stars \citep{bryson2020,dressing2015}. As the number of confirmed exoplanets increases, so too does the opportunity to characterize the composition and evolution of exoplanetary atmospheres. Atmospheric transmission spectroscopy with the \textit{James Webb Space Telescope} (JWST) is currently making the highest quality observations to date of exoplanet atmospheres. These observations are providing our first molecular inventories of gaseous exoplanets \citep{jwstco2} and are proving the first direct evidence for UV-driven photochemistry on planets beyond the solar system \citep{tsai2022}. As is now being shown by \textit{JWST}, the UV spectrum of a planet's host star drives photochemistry and photoionization in exoplanetary atmospheres, which in turn influences the overall composition of the atmosphere \citep{miguel2014,miguel2015,moses2011,venot2013}, the formation of photochemical hazes and aerosols \citep{he2018,kawashima2018,kawashima2019}, and powers atmospheric escape in both gaseous and terrestrial planets \citep{johnstone2015,vidalmadjar2004,murrayclay2009}. Thus, knowledge of the host star's UV spectrum is critical to the interpretation of current and future exoplanet atmosphere observations.

\subsection{Stellar UV effects on atmospheric chemistry and escape of Neptune to Jupiter sized planets}
\label{sec:intro_giant}
Spectral observations of molecules containing oxygen, nitrogen, and carbon are considered to be good candidates for potential biosignatures. Molecular reservoirs for these elements vary with pressure and temperature; in the upper atmospheres of hot planets with $700\text{ K}\gtrsim\text{T}\gtrsim1500\text{ K}$, oxygen, nitrogen, and carbon are typically contained in H$_2$O, N$_2$, and CO \citep{burrowsorton2010,madhusudhan2016,gao2021}. These molecules all have large wavelength-dependent photodissociation cross sections in the UV, typically peaking at less than 1300 $\text{\AA}$ \citep{loyd2016}. Thus, in the upper atmosphere of these planets, photochemical effects from UV irradiation can dominate the atmospheric composition, destroying NH$_3$, N$_2$, CO, CH$_4$, and H$_2$O, and leading to buildups of H, HCN, C$_2$H$_2$, N, and C \citep{moses2011,line2010,zahnle2009,miguel2014}. There is also evidence that increased UV radiation plays a role in creating the observed temperature inversion in hot-Jupiter atmospheres, where temperature increases with altitude.\citep{fu2022,lothringer2019,zahnle2009,knutson2010}.

The $>5000$ confirmed exoplanets so far exhibit a wide range of masses, from sub-Earth to tens of Jupiter masses, as well as orbital periods ranging from hours to tens of years. However, there is a notable dearth of Neptune-mass planets ($0.03\lesssim$ M/M$_{\text{J}}$ $\lesssim0.3$) with short orbital periods of $\lesssim5$ d, known as the ``Neptune desert" \citep{szab2011,mazeh2016}. \citet{owen2018} propose that a combination of high-eccentricity migration and photoevaporation can explain the presence of this desert. In order to quantify photoevaporation processes, we must understand the UV irradiation experienced by the exoplanet. Extreme-UV (EUV; $\sim100-912$ $\text{\AA}$) radiation, elevated by the small semi-major axes of these planets, is capable of heating the atmosphere to temperatures up to $\sim10^4$ K, driving thermal mass loss via hydrodynamic escape \citep{yelle2008,murrayclay2009,owenjackson2012,sanz-forcada2011}. On highly irradiated giant planets, the outflow may be sufficiently rapid that heavy elements are dragged along via collisions with hydrogen \citep{vidalmadjar2004,linsky2010,ballester2015,koskinen2013}. EUV driven hydrodynamic escape is particularly relevant for close-orbiting super-Earth to Neptune sized planets, potentially leading to complete evaporation of their gaseous envelopes on Gyr timescales, while Jupiter sized planets are more likely to retain their envelope over these timescales \citep{owenjackson2012,owenwu2016,fossati2017}.

\subsection{Stellar UV effects on atmospheric chemistry of sub-Neptune planets}
\label{sec:intro_terrestrial}
UV radiation also directly affects the observable features of sub-Neptune exoplanetary atmospheres. For example, atmospheric transmission spectra of GJ 1214b with the \textit{HST} Wide Field Camera 3 (WFC3) have revealed a ``flat" transmission spectrum from $\sim0.78-1.7$ $\mu$m; that is, the spectrum is missing the strong absorption features expected from H$_2$O and other molecules \citep{bean2010gj1214,berta2012gj1214}. \citet{kreidberg2014gj1214} concluded that the spectrum is inconsistent with a high molecular weight ($>50\%$ H$_2$O) atmosphere, arguing that the featureless spectrum is likely a result of optically thick clouds or photochemical hazes, whose formation is catalyzed by UV radiation. For sub- to super-Earth planets with tempertures $\lesssim700$ K, the dominant elemental reservoirs are H$_2$O, NH$_3$, and CH$_4$ \citep{burrowsorton2010,madhusudhan2016,fortney2021}. UV radiation from a planet's host star penetrates the upper atmosphere, photodissociating CH$_4$, and initiates photochemical reactions leading to the formation of opaque organic molecules which can cause features such as the observed flat transmission spectrum of GJ 1214b \citep{linsky2019,arney2017,miller-ricci2013}.

Remote sensing of exoplanet habitability relies heavily on the detection of gaseous biosignatures, particularly O$_2$, O$_3$, CO, and CH$_4$, but also including other hydrocarbons and N and S based gases (\citet{segura2005,desmarais2002}; see section 4 of \citet{schwieterman2018} for a more extensive overview of gaseous biosignatures). Molecular oxygen and ozone are readily detectable in the near-UV (NUV; $\sim1700-3200$ $\text{\AA}$), visible, and mid-IR and have the potential to be strong indicators of biologic activity. On Earth, O$_3$ is a byproduct of the photolysis of O$_2$, which is almost entirely sourced via oxygenic photosynthesis \citep{desmarais2002,kiang2007}. However, EUV and far-UV (FUV; $\sim912-1700$ $\text{\AA}$) radiation shortward of $\sim1700$ $\text{\AA}$---especially in the Lyman-$\alpha$ emission line, which can comprise $\sim37-75\%$ of the total FUV flux for M dwarfs \citep{france2013}---heavily influence oxygen chemistry, photodissociating CO$_2$ and H$_2$O and leading to escape of H and buildup of O and O$_2$ \citep{gao2015,hu2012,JPL}. The balance of oxygen and ozone in the atmosphere can be at least partially described by the Chapman mechanism \citep{chapman1930}, in which O$_2$ is photodissociated by FUV photons, which recombine to form O$_3$, and O$_3$ is in turn photodissociated by NUV and blue-optical photons, resulting in the production of O$_2$. Thus, the ratio of stellar FUV/NUV flux becomes critical for oxygen chemistry in the atmosphere; if a host star produces a large amount of FUV and relatively little NUV flux, a substantial O$_3$ atmosphere may arise entirely via photochemical processes \citep{segura2010,hu2012,tian2014,domagal2014,gao2015,schwieterman2018}.

\subsection{UV time variability of cool stars}
\label{sec:intro_variability}
Variability of UV radiation from a host star is critical to photochemistry and atmospheric stability of exoplanets. Solar EUV flux varies by factors up to $\sim100$ on minute timescales during intense flares \citep{woods2012}. In G-, K-, and M stars, quiescent FUV radiation is emission line dominated but continuum emission can become the dominant UV luminosity source during stellar flares \citep{kowalski2010,loyd2018}. M and K dwarfs in particular exhibit regular flare activity, even in old and inactive stars, and the energy released during these events may account for more FUV flux than the quiescent emission over stellar lifetimes \citep{loyd2018,france2020}. Knowledge of stellar flare rates and energies are therefore necessary to allow estimates of lifetime-integrated UV flux experienced by exoplanets, especially those being assessed for their potential habitability. Solar observations have shown that many high-energy flaring events are associated with an accompanying coronal mass ejection (CME) and that larger flare fluxes result in larger CME masses \citep{munro1979,aarnio2011}. These CMEs result in highly energetic accelerated particles which impact planetary atmospheres, significantly enhancing pickup ions and leading to dramatic increases in atmospheric escape rate \citep{jakosky2015,lammer2006coronal,airapetian2017}. Furthermore, it has been shown that energetic particle deposition into the atmospheres of terrestrial planets can lead to significant changes in observable atmospheric oxygen abundances \citep{segura2010,tilley2019}.

\subsection{The MUSCLES and Mega-MUSCLES Treasury Surveys}
\label{sec:intro_MUSCLES}
With the growth of exoplanetary science and the associated awareness of the importance of the host star's UV radiation field on the evolution of exoplanetary atmospheres, the community resources devoted to UV characterization of cool stars have increased \citep{france2018,HAZMAT1,FUMES3}. This is particularly important since it has been shown that empirical scaling relations alone are insufficient to model the evolution of planetary atmospheres, and the extended UV continuum and UV emission lines are necessary to generate accurate models \citep{teal2022,peacock2022}. The MUSCLES Treasury Survey (\textit{HST} Cycle 22; PI---France) began to address this dearth of observations by creating panchromatic 5 $\text{\AA}$-5.5 $\mu$m SEDs of M and K dwarfs which have since been used extensively to study the importance of the UV radiation environment on exoplanets \citep[for example,][]{kawashima2018,lora2018,chen2021}. The MUSCLES SEDs consist of observational spectra in the X-ray (5-50 $\text{\AA}$: \textit{XMM-Newton} and \textit{Chandra X-ray Observatory}) and UV (1170-5700 $\text{\AA}$: \textit{HST}), empirical estimates of the EUV (100-1170 $\text{\AA}$) which cannot be presently observed owing to lack of an operating EUV observational facility \citep{france2022escape}, and stellar atmospheric models of the IR (5700 $\text{\AA}$-5.5 $\mu$m) \citep{loyd2016}. Stars in the MUSCLES survey covered a range of spectral types from K1V - M5V and all but one were considered ``optically inactive'' due to having H$\alpha$ in absorption. Despite this classification, all showed chromospheric and coronal activity \citep{france2016,youngblood2016,loyd2016}.

The MUSCLES survey was subsequently expanded to include an additional 13 M dwarfs in the Mega-MUSCLES survey (\textit{HST} Cycle 25; PI---Froning). Mega-MUSCLES has a particular focus on low-mass (M$<0.3$ M$_\sun$) stars with a range of spectral types from M0-M8, including Barnard's star and TRAPPIST-1. Mega-MUSCLES observations revealed a number of UV and X-ray flares on stars with a range of activity levels \citep{froning2019,france2020}, demonstrating that optical activity indicators are poor predictors of the high energy variability of cool stars and further reinforcing the need for direct UV and X-ray observations of specific planet hosting stars.

\subsection{The MUSCLES Extension for Atmospheric Spectroscopy}
\label{sec:intro_MEATS}
The recent launch of \textit{JWST} has begun an unprecedented era in exoplanet atmospheric characterization. Atmospheric spectroscopy of exoplanets ranging from Earth-sized terrestrial planets through giant hot-Jupiters are currently being obtained via the \textit{JWST} Early Release Science (ERS) and Guaranteed Time Observations (GTO) programs. As described above, the high-energy SED of the host stars will be crucial to accurately interpret the results of these spectroscopic observations. We have identified 11 \textit{JWST} guaranteed time targets which have no previous UV observations in the \textit{HST} archive. In this work, we present the MUSCLES Extension for Atmospheric Transmission Spectroscopy, which extends the original MUSCLES survey over a larger range in stellar mass to include these 11 previously uncharacterized stars. The stars in this work range from M6-F4 and host planets ranging from super-Earths to hot-Jupiters. We expand on the methods of the MUSCLES survey and create panchromatic SEDs of these stars, characterize them in relation to other known planet and non-planet hosting stars, and address selective bias towards observing low activity stars which may impact our interpretations of observed exoplanet atmospheres.

We structure the paper as follows: Section \ref{sec:obs} describes the observational campaign, including \textit{HST}, \textit{Chandra}, and \textit{XMM-Newton} observations, and the methods used to reconstruct the currently unobservable regions in the EUV. In section \ref{sec:results} we show the results of our observations and put them in context relative to the broader population of stellar surveys. We also discuss the importance of studying the time variability of host stars, motivated by the detection of two X-ray flares in the star L 98-59. Finally, we summarize the main results of this work in Section \ref{sec:summary}. A list of observed targets including particular observation details, SED construction, and descriptions of planetary systems can be found in Appendix \ref{sec:targets}. Lists of UV emission line fluxes can be found in Appendix \ref{sec:emission_fluxes}.

\section{Observations} \label{sec:obs}

Observational data were obtained from \textit{HST}, \textit{Chandra}, and \textit{XMM-Newton}. All \textit{HST} observations were obtained through a dedicated observing program for this survey (\textit{HST} Cycle 28, program ID 16166; PI---France), while \textit{Chandra} and \textit{XMM-Newton} were a combination of new and archived observations. In Table \ref{tab:targets} we present a brief overview of each target in the MUSCLES Extension in order of decreasing effective temperature. For a detailed description of the planetary systems and their relation to the \textit{JWST} ERS and GTO programs, as well as X-ray and UV data quality, major emission characteristics, and details of X-ray/FUV proxy stars (see Section \ref{sec:FUVproxy}), we refer the reader to Appendix \ref{sec:targets}.

\begin{deluxetable}{ccCCCCC}[h] 
    \tablecaption{List of Targets\label{tab:targets}}
    \tablehead{
    \colhead{Star}  &   \colhead{Sp.Type}   &   \colhead{T$_{\text{eff}}$ [K]} &   \colhead{Distance [pc]} &   \colhead{Mass [M$_\sun$]}   &   \colhead{Radius [R$_\sun$]} &   \colhead{\# of planets\tablenotemark{a}}
    }
    \startdata
    WASP-17 &  F4   &   6548    &   405.0^{+8.8}_{-8.4} &   1.57\pm0.092   &   1.31\pm0.03   &   1\\
    HD 149026  &   G0   &   6084    &   75.0\pm1.7  &   1.46\pm0.08 &   1.34\pm0.02 &   1\\ 
    WASP-127    &   G5  &   5828    &   159.0\pm1.2 &   1.31\pm0.05 &   1.33\pm0.03 &   1\\
    WASP-77A    &   G8  &   5605  &   105.2\pm1.2   &   1.00\pm0.05   &   0.96\pm0.02   &   1\\
    TOI-193 &   G7  &   5443    &   80.4\pm0.3  &   1.02^{+0.02}_{-0.03}    &   0.95\pm0.01   &   1\\
    HAT-P-26    &   K1  &   5062    &   141.8^{+1.2}_{-1.1} &   0.82\pm0.03 &   0.79^{+0.10}_{-0.04}    &   1\\
    HAT-P-12    &   K5  &   4653    &   142.8\pm0.5 &   0.73\pm0.02   &   0.70^{+0.02}_{-0.01} &   1\\
    WASP-43 &   K7  &   4124    &   86.7\pm0.3  &   0.72\pm0.03  &   0.67\pm0.01  &   1\\
    L 678-39    &   M2.5    &   3490    &   9.4\pm0.01   &   0.34\pm0.01  &   0.34\pm0.02   &   3\\
    L 98-59 &   M3  &   3429   &   10.6\pm0.003  &   0.31\pm0.01    &   0.31\pm0.01   &   4\\
    LP 791-18   &   M6  &   2949    &   26.5\pm0.06  &   0.14\pm0.01   &   0.17\pm0.02   &   3\\
    \enddata
    \tablenotetext{a}{Sourced from NASA Exoplanet Archive. We report only the number of confirmed exoplanets.}
    References: From top to bottom, (1) \citet{anderson2010}; (2) \citet{sato2005}; (3) \citet{lam2017}; (4) \citet{maxted2013}; (5) \citet{jenkins2020}; (6) \citet{hartman2011}; (7) \citet{hartman2009}; (8) \citet{hellier2011}; (9) \citet{lugue2019}; (10) \citet{kostov2019}; (11) \citet{crossfield2019}. References represent the announcement of discovery of the exoplanet(s). Further references are listed in the more detailed system descriptions in Appendix \ref{sec:targets}.
\end{deluxetable}

\subsection{FUV and NUV}\label{sec:FUV_and_NUV}

Unless otherwise stated, we refer to FUV as $912\text{ \AA} < \lambda < 1700\text{ \AA}$ and NUV as $1700\text{ \AA} < \lambda < 3200\text{ \AA}$. We employed the STIS G140L and G230L gratings for the FUV and NUV continuum and emission lines, respectively, and the G140M grating for higher resolution spectra in the Lyman-$\alpha$ emission region. Finally, in order to calibrate the UV data to visible/IR photospheric models and ground-based spectra, we obtained optical observations using the STIS G430L grating. The exposure times were estimated based on the minimum amount of time required to achieve signal-to-noise (S/N) of $\text{S/N}\simeq10$ per resolution element in the characteristic line and continuum regions: Ly$\alpha$ (STIS G140M, $\lambda=1216.0$ $\text{\AA}$), C II (STIS G140L), and NUV continuum (STIS G230L, $\lambda=2820$ $\text{\AA}$, longward of Mg II), and $\text{S/N} > 20$ in the optical regions. We create simulated spectra for exposure time estimates by taking stars of similar spectral types from the HST archive (Procyon, $\alpha$ Cen A, $\epsilon$ Eri, HD 85512, and GJ 832), scaled to the V magnitude of the target star and use the STScI exposure time calculator\footnote{\url{https://etc.stsci.edu/etc/input/stis/spectroscopic/}} to estimate the minimum exposure time. As we shall discuss in Section \ref{sec:results}, the MUSCLES Extension targets were lower activity than expected and the S/N in the FUV was lower than that calculated in the observation proposal. In the NUV/optical, the S/N calculations were accurate with $S/N > 20$ per resolution element. In total, our \textit{HST} spectra typically span $\sim1150-5500$ $\text{\AA}$.

For some cool stars of spectral type K5 and later, the required exposure time to obtain an adequate S/N was prohibitively long and we opted to exclude G140L and G140M observations. We were thus unable to acquire direct FUV emission line measurements for these targets. This includes HAT-P-12, LP 791-18, and WASP-43.

L 678-39 exceeded the STIS G140L bright object protection limits, and we thus utilized the \textit{HST} Cosmic Origins Spectrograph (COS) G130M (selecting the $\lambda$1222 CENWAVE to avoid specific bright emission lines) and G160M modes, rather than STIS. For similar reasons, we have observed HD 149026 with STIS E230M rather than G230L.

We reduced the \textit{HST} data as follows:

First, we examined all observations for pointing or data quality errors. Several observations (G140M: HD 149026, L 98-59, TOI-193; G140L: WASP-17; G230L: LP 791-18) had incorrect extraction regions during the X1DCORR routine, presumably due to low signal. We examined the flat fielded images to determine proper extraction regions and re-extracted them manually using the Python package stistools\footnote{\url{https://stistools.readthedocs.io/en/latest/}}. We also found poor data quality flags on the blue end of the G430L CCD for all observations, typically spanning $\sim100$ pixels; these were mostly pixels that were flagged for having zero flux (ie: flag 16384 in the data quality array; see section 2.5 of \citeauthor{STISdata} \citeyear{STISdata}). The FUV MAMA detectors did not show any serious data quality issues.

After screening the observations and removing portions with poor data quality, we proceeded to coadd the spectra for any target that had multiple observations with the same grating. We first interpolated each spectrum onto a common wavelength grid with $\Delta\lambda=0.5$ $\text{\AA}$, oversampling the native resolution of the gratings. This process conserves the total observed flux while allowing us to perform a simple coaddition. We then performed a coaddition of the spectra using an exposure time weighted average.

After performing the coaddition for each grating observation, we examined the final S/N of each and culled data that we considered to be of poor quality; we chose a threshold of S/N$>3$ per pixel for data to be usable in the final spectrum. This resulted in all of the continua of all G140L observations being too low S/N to include in the final spectra, although we do find emission lines above the S/N threshold for most targets. G230L spectra were typically low S/N on the blue end until $\sim2000-2300$ $\text{\AA}$, depending on effective temperature, at which point the photospheric emission begins to pick up. G430L spectra were all above the S/N threshold after culling the poor data quality pixels on the blue end of the detector. Finally, we compare the reported stellar B-V color to a table of unreddened colors of the appropriate spectral type to determine if a dereddening procedure is required to account for interstellar extinction. We found A$_v < 0.1$ for all targets and thus do not account for reddening in any of the analysis.

After cleaning and coadding the spectra, we measured the emission line fluxes of the seven lines listed in Table \ref{tab:fluxes}. 

For emission lines with S/N greater than the threshold, the reported flux is the numerically integrated flux over the continuum-subtracted line region:

\begin{equation}
    F_{ion} = \int^{\lambda_0+\delta\lambda}_{\lambda_0-\delta\lambda}F(\lambda) d\lambda
\end{equation}

\noindent where $F(\lambda)$ is the continuum subtracted flux density. The continuum flux density is estimated from a polynomial fit to the continuum on either side of the emission line. The integration width, $\delta\lambda$, was selected by hand for each line to accommodate varying line widths and was typically 2.5 $\text{\AA}$ for the low resolution G140L spectra.

For targets with emission line fluxes below the S/N threshold, we report the RMS value of the flux density over the continuum-subtracted line region as an upper limit on the emission line flux.

\subsection{X-ray data}
\label{sec:xray}
The MUSCLES Extension included X-ray data of 4 stars via \textit{Chandra} observations and 7 via \textit{XMM-Newton} observations. The data were a combination of new observations obtained for this program as well as archival observations. The new \textit{XMM-Newton} observations of WASP-43 and L 98-59 were taken concurrently with the UV observations obtained in this work. Source X-ray spectra were extracted from a circular region with a 2.5" radius around the proper motion corrected source location and an annular background region centered on the target location encompassing as much area as possible without including other sources. Some targets were close to the edge of the detector chip and a background region centered on the target would extend beyond the edges of the chip; in these cases, we used a circular background region from a nearby representative area. \textit{Chandra} data were analyzed using the Chandra Interactive Analysis of Observations \citep[CIAO;][]{ciao2006} software. We use the CIAO \textit{dmlist} routine to obtain background subtracted count rates. For \textit{XMM-Newton} observations, we use the Scientific Analysis System (SAS 20.0.0; \citet{SAS}) with the standard procedure and filters. Photon events were limited to those with an energy range of 0.3-10 keV to remove spurious high-energy particle events. Each target was screened for both source and background flaring. Background flares are common in \textit{XMM-Newton} observations and were removed from several targets. L 98-59 is the only target which showed a source flaring event. Flares occurred in both observations of L 98-59 which continued throughout the duration of each exposure. These flares are discussed in detail in Section \ref{sec:flare}.

Five targets were successfully detected at the 3$\sigma$ level; however, the number of counts remaining in the source regions were insufficient to allow robust spectral modeling. Three of four observations with the \textit{Chandra} ACIS-S detector and three of eight observations with the \textit{XMM-Newton} EPIC pn detector were non-detections at the $3\sigma$ level. To estimate a flux for each target we use the PIMMS\footnote{The Portable Interactive Multi-Mission Simulator, http:/cxc/harvard.edu/toolkit/pimms/jsp} tool, assuming a thermal plasma model with a temperature of 0.43 keV
and a hydrogen column density estimated based on stellar distance using the relations from \citet{wood2005}. For detected sources we input the observed count rate. For non-detected sources we input a theoretical count rate which would produce a $3\sigma$ detection given our exposure time and report the estimated flux as an upper limit.

\subsection{X-ray and FUV proxy spectra}\label{sec:FUVproxy}
Due to the amount of X-ray non-detections and the low S/N of our G140L observations, we opted to use scaled proxy spectra to represent the X-ray ($\sim5-100\text{ \AA}$) and FUV continuum. Proxy stars were chosen to have similar spectral type, age, and activity levels, based on effective temperature, rotation period, and $\log{R'_{HK}}$. Spectra for the proxy stars were obtained from the publicly available MUSCLES \citep{france2016} archives. The MUSCLES spectra used the \textit{HST} COS to obtain UV measurements down to $\sim1150$ $\text{\AA}$, and a combination of \textit{XMM-Newton} and \textit{Chandra} observations and Astrophysical Plasma Emission Code \citep[APEC;][]{apec} models for the X-rays. After selecting a suitable proxy star, the proxy spectrum was scaled to the blue end of the NUV spectrum (G230L) via a least-squares fitting method minimizing the quantity$((F_{ref}-\alpha\times F_{proxy})/\sigma_{ref})^2$ where $F_{ref}$ represents the G230L flux of the MUSCLES Extension target, $F_{proxy}$ the flux of the proxy spectrum, $\alpha$ the non wavelength-dependent scaling factor applied to the proxy spectrum, and $\sigma_{ref}$ the 1$\sigma$ error of the observed G230L spectrum. We applied this routine to a region approximately $100$ $\text{\AA}$ wide where the proxy spectrum overlaps with the blue end of the observed G230L spectrum. The continua of the scaled proxy spectra match the continua of the observed STIS spectra within the $1\sigma$ uncertainty of the STIS data. Considering the importance of UV emission lines in atmospheric modeling, we replace the emission lines of the proxy spectra with the measured emission lines from our MUSCLES Extension observations where possible; this provides the best balance between a representative continuum and ground-truth emission line fluxes. In regions where our observed emission lines exceed the S/N$>3$ per pixel threshold, we remove the proxy spectrum and replace it directly with our observed data. Regions which include the observed emission lines are recorded in the instrument data column of the final SED data products, represented by a bit value corresponding to the instrument and grating with which the line was measured (Section \ref{sec:panchrom}). For lines which are below the S/N threshold, we first construct a representative line by assuming a Gaussian emission with a width of 60 km/s for M stars and 70 km/s for all others, based on \citet{france2020} and \citet{france2010}, and match the total integrated flux. We then replace the emission line region in the proxy spectrum with the constructed flux-matched emission line. Proxy spectra are used in the regions of 5-100 $\text{\AA}$ and again from 1170 $\text{\AA}$ to the beginning of the high S/N NUV observations, which ranges from 1750-2600 $\text{\AA}$. The FUV proxy region excludes the reconstructed Lyman-$\alpha$ range from $\sim$1212-1220 $\text{\AA}$.

\subsection{Lyman-$\alpha$}
\label{sec:lya}
The Lyman-$\alpha$ emission line is heavily attenuated by neutral hydrogen in the ISM, with the core of the line often being unobservable for even the nearest stars, leaving only the wings of the line observed. Lyman-$\alpha$ flux plays an important role in the photochemistry of exoplanet atmospheres and therefore it is crucial to properly reconstruct the intrinsic line profile in order to accurately model atmospheric chemistry \citep{miguel2015,arney2017}. We follow the approach of \citet{youngblood2022} to reconstruct the intrinsic Lyman-$\alpha$ profile from the ISM-attenuated, observed spectra of the stars with good S/N G140M data. We simultaneously fit a model of the intrinsic profile and the ISM absorption to the observed line wings. The functional form of the intrinsic line profile is a Voigt emission profile with a self-reversal that follows the shape of the emission:

\begin{equation}\label{eq:voigt}
    F_{emission}^\lambda = \mathcal{V}^\lambda\cdot\exp{(-p\mathcal{V}_{norm}^\lambda)}
\end{equation}

\noindent where $\mathcal{V}^\lambda$ is the Voigt emission profile, $p$ the self-reversal parameter, and $\mathcal{V}_{norm}^\lambda$ the peak-normalized Voigt profile. The self-reversal parameter is allowed to vary between 0 and 3. Larger values of $p$ result in a deeper self-reversal profile.

The ISM absorption is modeled as two Voigt profiles without self-reversal parameters; one for hydrogen and one for deuterium.

We use the reconstructed line for the region of $\sim1212-1220$ $\text{\AA}$ in the final SEDs. For targets where G140M observations were not feasible or the S/N was not sufficient to fit with the MCMC method, we estimated the total Lyman-$\alpha$ flux by using a power-law relation to the total Mg II $\lambda\lambda2799,2803\text{ \AA}$ surface flux \citep{wood2005,youngblood2016}. After estimating the Lyman-$\alpha$ flux based on the Mg II relation, we create a line profile of the form given by equation (\ref{eq:voigt}) by selecting a fixed self-reversal parameter and iterating through a range of amplitudes until the integrated flux of the self-reversed Voigt profile matches the flux estimated by the Mg II relation. The self-reversal parameter was chosen to be 1.5 for M type stars, 2.0 for K, and 2.4 for G and F types based on the results from \citep{youngblood2022}. The integrated Lyman-$\alpha$ flux of the Voigt profiles match the estimated flux to within 0.05\%.

\begin{figure}[h]
    \centering
    \includegraphics[width=0.8\textwidth]{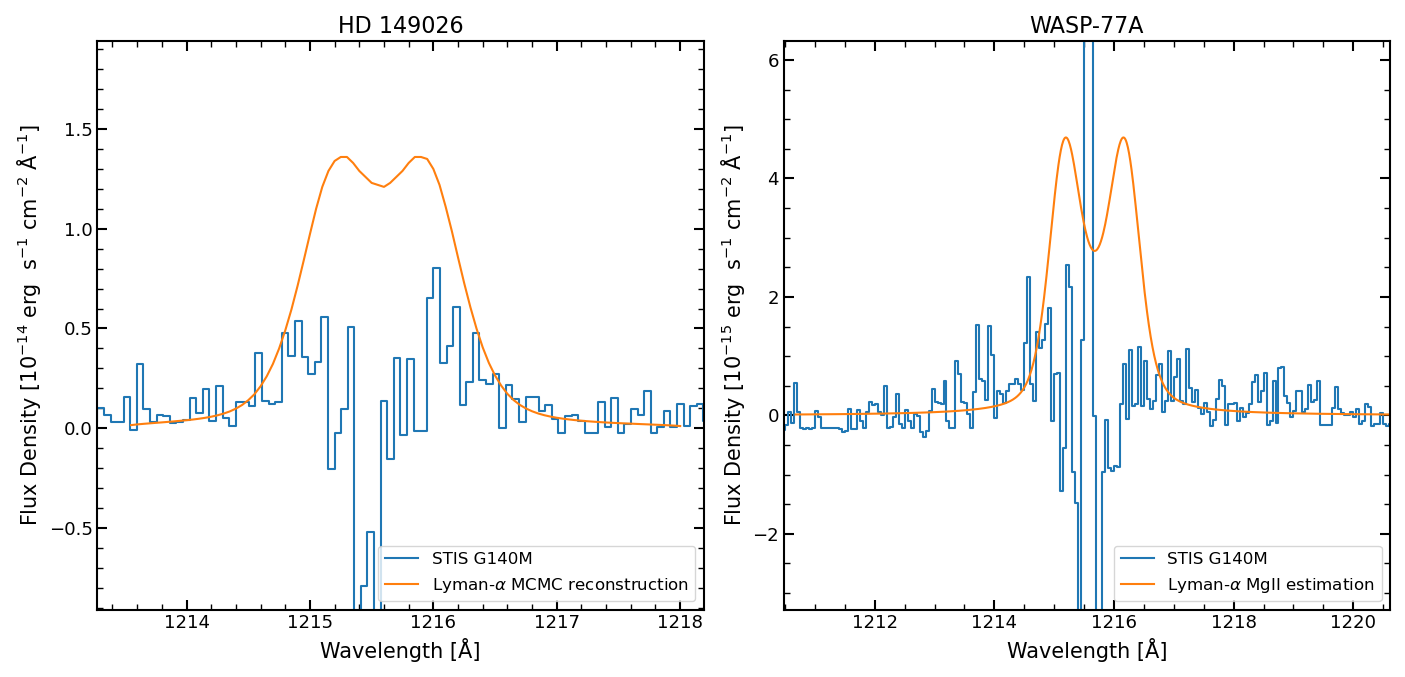}
    \caption{Left: Lyman-$\alpha$ reconstruction for HD 149026 using the \citet{youngblood2022} MCMC method. Right: Lyman-$\alpha$ reconstruction for WASP-77A using the Mg II flux estimation method.}
    \label{fig:lya-recon}
\end{figure}

Figure \ref{fig:lya-recon} shows a reconstructed profile for two stars; HD 149026, for which the G140M observation was sufficient to fit with the MCMC method, and WASP-77A, for which we used the Mg II flux estimation. The WASP-77A spectrum exhibits both large negative fluxes and spurious peak towards the line center; this is a result of poor background subtraction of geocoronal Lyman-$\alpha$.

\subsection{Visible through IR}
\label{sec:VisIR}
\begin{figure}[h]
    \centering
    \includegraphics[width=0.8\textwidth]{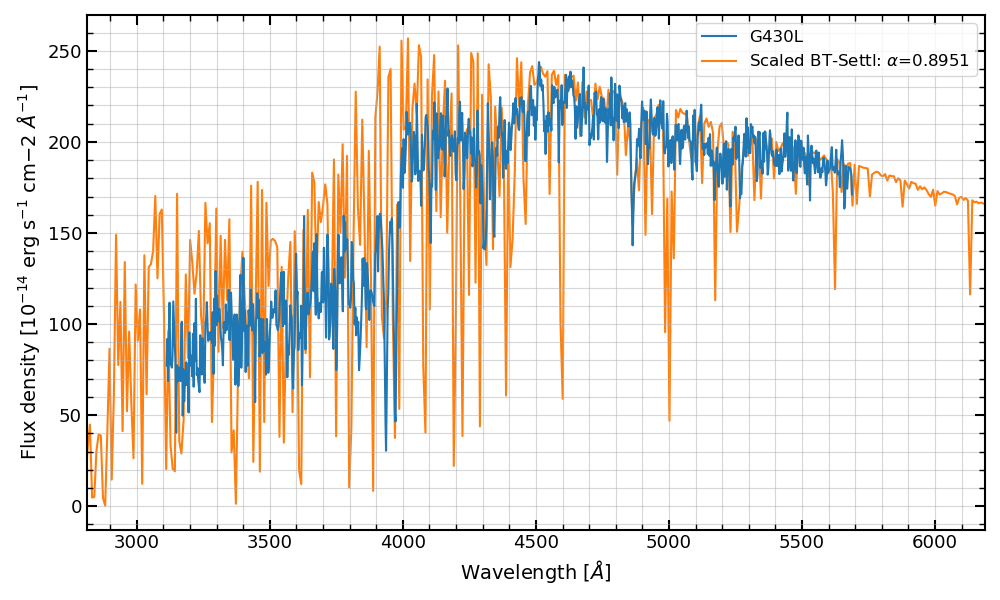}
    \caption{Result of the HD 149026 BT-Settl scaling routine plotted with the STIS G430L spectrum. The BT-Settl stellar atmosphere model is convolved to G430L resolution and scaled by a factor of $\alpha=0.8951$.}
    \label{fig:BT-Settl}
\end{figure}

The G430L spectra extend into visible wavelengths up to 5700 $\text{\AA}$. After this point, we use the BT-Settl  stellar atmosphere models \citep{allard2011} to extend the SEDs into the IR up to $5.5\mu$m. The BT-Settl models cover a grid of effective temperatures (T$_{\text{eff}}$) and surface gravity ($\log{g}$) and provide a flux observed at Earth based on stellar radius and distance. We scale the BT-Settl model to the NUV spectra from G430L as follows:

First, we convolve the high resolution BT-Settl model to match the resolution of the G430L data. We then take region $>5000$ $\text{\AA}$ where the BT-Settl model overlaps with the G430L spectrum and scale the model using the same least-squares method described in Section \ref{sec:FUVproxy}. The fits were examined by eye and found to be well representative of the shape of the observed spectra. Figure \ref{fig:BT-Settl} shows the BT-Settl atmospheric model for $6000\text{ K} < \text{T}_{\text{eff}} < 6100\text{ K}$ and $4.0 < \log{g} < 4.5$ scaled to the G430L spectrum of the G0 star HD 149026 using this method.

\subsection{EUV estimations}
\label{sec:EUV}
The extreme ultraviolet is heavily attenuated by the interstellar medium, particularly in the Lyman continuum region from the photoionization point of H at 912 $\text{\AA}$ down to $\sim400$ $\text{\AA}$. Interstellar extinction, combined with the lack of an operating EUV observatory, means we currently do not have the ability to observe the EUV spectra of stars other than the Sun, and the \textit{HST} sensitivity drops sharply below $\sim1170$ $\text{\AA}$; we therefore present estimations of the EUV flux from 100-1170 $\text{\AA}$. The EUV spectrum is estimated in nine  bandpasses from 912-1170 $\text{\AA}$ using empirically derived relations between total Lyman-$\alpha$ flux and EUV flux from \citet{linsky2014}. \citet{linsky2014} used a combination of solar models ($\lambda<2000$ $\text{\AA}$) and direct observations, utilizing the \textit{Far Ultraviolet Spectroscopic Explorer} (\textit{FUSE}; 912-1170 $\text{\AA}$) and the \textit{Extreme Ultraviolet Explorer} (\textit{EUVE}; 100-400 $\text{\AA}$), to show that the ratio of EUV to Ly$\alpha$ flux varies slowly with the total Ly$\alpha$ flux. They established direct ratios of $F(\Delta\lambda)/F(\text{Ly}\alpha)$, where $\Delta\lambda$ is the wavelength region of the $\sim100$ $\text{\AA}$ bandpass, and both $F(\Delta\lambda)$ and $F(\text{Ly}\alpha)$ are scaled to 1 AU. After calculating the EUV flux in the bandpass we re-scale the flux to the appropriate stellar distance.

\subsection{Panchromatic Spectrum Assembly}
\label{sec:panchrom}
With all of the data products described above in hand we developed a procedure for stitching the spectra together into a continuous panchromatic spectrum. We first define a prioritization order for the spectral segments, placing the most priority on direct observations with good S/N, followed by Lyman-$\alpha$ reconstructions and EUV estimations, and finally scaled proxy and BT-Settl spectra.

Keeping in mind our goal of using direct observations wherever possible, we opted not to perform any optimization routine to determine the location of the join. Instead, we directly inserted whichever data product was highest on the priority list for the given wavelength range. This occasionally results in a small jump discontinuity at the joining location. We investigated these discontinuities by fitting a smooth spline between the two regions and finding the total amount of ``missing" flux, which was less than 0.7\% of the flux over the same region without including a smooth join. We therefore consider the discontinuities to be negligible and do not attempt to correct them. In cases where two direct observations overlap, we give priority to whichever observation has higher S/N over the region in question.

\begin{figure}[h]
    \centering
    \includegraphics[width=\textwidth]{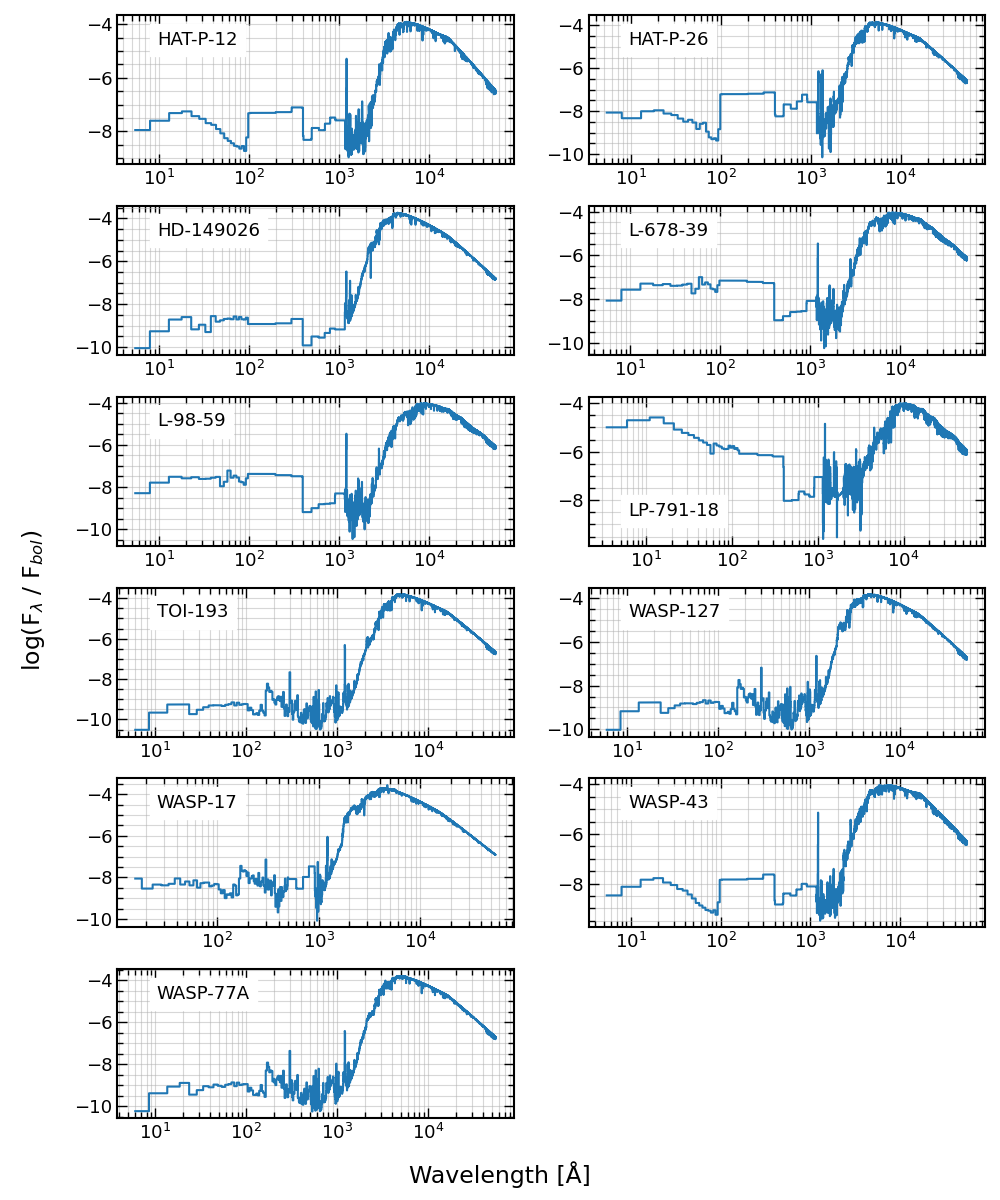}
    \caption{Panchromatic stellar SEDs from 5 $\text{\AA}$-5.5 $\mu$m. The vertical axis represents the flux density normalized by the bolometric flux, spanning approximately $10^{-10}-10^{-4}\text{ \AA}^{-1}$. All spectra have been binned to 5 $\text{\AA}$ per pixel for visualization.}
    \label{fig:panchrom}
\end{figure}

Figure \ref{fig:panchrom} shows all of the final panchromatic SEDs.

The MUSCLES Extension SEDs are available as high-level science products on the MUSCLES portal hosted on the MAST archive\footnote{\url{https://archive.stsci.edu/prepds/muscles/}} as a FITS file containing a PrimaryHDU with general observation information and a BinTableHDU containing the spectral data. We provide two versions of each SED: one which retains the native instrument or model resolutions, and one which is rebinned to a constant 1 $\text{\AA}$ resolution. Each panchromatic SED provides the following information:

\begin{itemize}
    \item Bin: Midpoint and edges of the wavelength bins in [$\text{\AA}$]. 
    \item Flux density: Measurement and error of the flux density in [erg s$^{-1}$ cm$^{-2}$ $\text{\AA}^{-1}$] as well as the value normalized by the bolometric flux [$\text{\AA}^{-1}$]
    \item Exposure: MJD of the beginning of the first contributing exposure, the end of the last contributing exposure, and the cumulative exposure time [s]
    \item Normalization: any normalization factor applied to flux bin
    \item Instrument: a bit-wise flag identifying the source of the flux data for the bin. Note that for binned spectra, we may have combined adjacent bins from different sources. This is accounted for in the bit-wise instrument flag by adding the bit value for each of the respective instruments.
\end{itemize}

\section{Results}
\label{sec:results}
\citet{france2018} showed that exoplanet hosting stars selected from RV and transit methods exhibit lower UV activity than the general field population. In order to further understand the potential effects of selecting stars chosen from planet detection surveys to use as templates for atmospheric modeling we present a comparison of our targets to previously studied planet hosting and ``non-planet hosting" stars. We examine first the fractional X-ray luminosity (Section \ref{sec:xray_analysis}), which is an indicator of coronal activity levels, then the FUV/NUV flux ratio and fractional UV emission line luminosities (Sections \ref{sec:FUV_NUV_ratio} and \ref{sec:ion_ratios}), which are indicative of chromospheric/transition-region activity levels, and finally the UV flux environment experienced by the planets orbiting our target stars compared to those in the MUSCLES survey (Section \ref{sec:planet_flux}). The X-ray flaring events of L 98-59 are discussed in Section \ref{sec:flare}.

\subsection{X-ray flux}
\label{sec:xray_analysis}
\begin{figure}[h]
    \centering
    \includegraphics[width=0.8\textwidth]{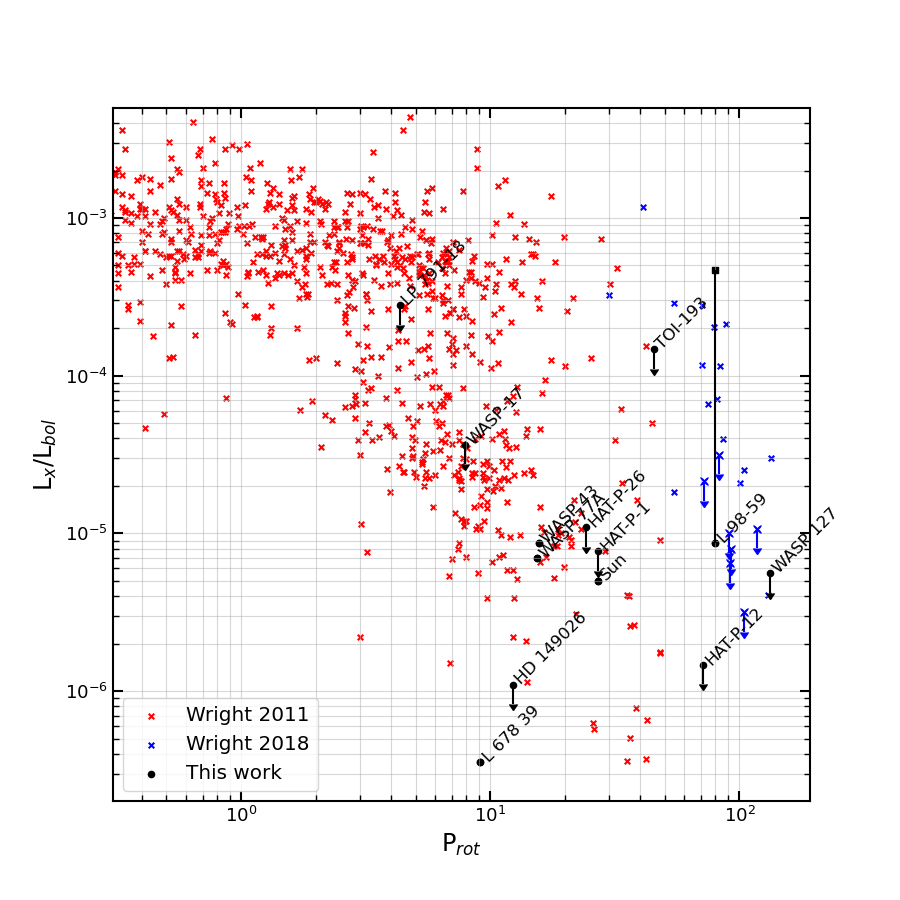}
    \caption{Fraction of bolometric luminosity emitted in the X-ray. \citet{wright2011} represents solar and late-type stars. \citet{wright2018} represents fully convective M dwarfs. The black square connected by vertical line to L 98-59 represents the X-ray luminosity during only the flaring event. The black circle representing L 98-59 shows the quiescent X-ray luminosity.}
    \label{fig:X-ray}
\end{figure}

The fraction of stellar bolometric flux emitted in the X-ray band has been shown to be correlated with chromospheric and coronal activity indicators such as rotation period, $R'_{HK}$, and $R_{H\alpha}$, and thus is a useful measure of stellar activity levels \citep{kostov2019,katsova2011,he2019,linsky2020}. Figure \ref{fig:X-ray} shows the fractional X-ray luminosity of MUSCLES Extension stars compared to a large survey by \citet{wright2011,wright2018}. \citet{wright2011} observed 824 solar and late-type stars to study the relation between rotation period and stellar activity; they extended this survey in \citet{wright2018} to include a sample of 19 fully convective M dwarfs.

Here we use the X-ray flux or upper limits from PIMMS as described in section \ref{sec:xray} and define the bolometric flux based on effective temperature:

\begin{equation}
    F_{bol} = \sigma T_{\text{eff}}^4\left( \frac{R_*}{d}\right)^2
\end{equation}

\noindent where $\sigma$ is the Stefan-Boltzman constant, $R_*$ the stellar radius, and $d$ the stellar distance. Using this definition of bolometric flux allows for consistency between measurements of our own targets as well as to those of \citet{france2018}, who use the same definition. Results from the PIMMS analysis are listed in Tables \ref{tab:xray_XMM} and \ref{tab:xray_chandra}.

\begin{deluxetable}{cccCCCCCc}[h]
    \tablewidth{\linewidth}
    \tablecaption{\textit{XMM-Newton} Observations\label{tab:xray_XMM}}
    \tabletypesize{\scriptsize}
    \tablehead{
    \colhead{Star}  &   \colhead{Obsid} &   \colhead{Exp time [ks]}    &   \colhead{Net counts} &   \colhead{Count rate [s$^{-1}$]\tablenotemark{a}} &   \colhead{F$_x$ [erg s$^{-1}$ cm$^{-2}$]} &   \colhead{L$_x$ [erg s$^{-1}$]}  &   \colhead{L$_x$/L$_{bol}$} &   \colhead{Ref}
    }
    \startdata
    HAT-P-12    &   0853380901   &   2.7875  &   1   &   <3.23\times10^{-3} &   <3.59\times10^{-15}    &   <8.79\times10^{27}   &   <1.10\times10^{-5}  &   1\\
    HAT-P-26    &   0804790101   &   14.041  &   7  &   <6.41\times10^{-4}  &   <7.62\times10^{-16}  &   <1.84\times10^{27}   &   <1.10\times10^{-6}    &   2\\
    HD 149026   &   0763460301   &   15.55  &   62  &   2.32\pm0.51\times10^{-3}  &   5.19\pm1.14\times10^{-15}  &   3.59\pm0.79\times10^{27}   &   3.56\pm0.78\times10^{-7}    &   3\\
    L 678-39    &   0840841501   &   22.409  &   69  &   2.67\pm0.43\times10^{-3}  &   2.73\pm0.44\times10^{-15}  &   2.65\pm0.43\times10^{25}   &   3.99\pm0.64\times10^{-7}   &   4\\
    WASP-43 &   0871800101   &   21.21  &   81 &   3.39\pm0.43\times10^{-3}  &   3.56\pm0.45\times10^{-15}  &   3.22\pm0.40\times10^{27}   &   5.62\pm0.70\times10^{-6} &   5\\
    WASP-127    &   0853380601   &   1.0274  &   0   &   <8.76\times10^{-3} &   <1.26\times10^{-14} &   <3.86\times10^{28}  &   <5.02\times10^{-6}  &   1\\
    L 98-59 &   0871800201  &   2.896   &   36  &   1.00\pm0.20\times10^{-2}  &   3.20\pm0.06\times10^{-14}  &   3.98\pm0.73\times10^{26} &   8.72\pm0.17\times10^{-6}  &   5\\
    L 98-59 &   0871800301  &   2.896    &   49  &   1.20\pm0.20\times10^{-2}    &   3.50\pm0.05\times10^{-14}   &   3.98\pm0.55\times10^{26}    &   9.43\pm0.13\times10^{-6}   &   5\\
    \enddata
    \tablenotetext{a}{Upper limits represent the count rate required to produce a 3$\sigma$ detection given the listed exposure time}
    References: (1) XMM-Newton Target of Opportunity (proposal ID 085338; PI Shartel), (2) \citet{sanz-forcadaXMMprop}, (3) \citet{salzXMMprop}, (4) \citet{stelzerXMMprop}, (5) \citet{meatsXrayprop}
\end{deluxetable}
    
\begin{deluxetable}{cccCCCCCc}[h]
    \tablecaption{\textit{Chandra} Observations\label{tab:xray_chandra}}
    \tabletypesize{\scriptsize}
    \tablehead{
    \colhead{Star}  &   \colhead{Obsid} &   \colhead{Exp time [ks]}    &   \colhead{Net counts} &   \colhead{Count rate [s$^{-1}$]\tablenotemark{a}} &   \colhead{F$_x$ [erg s$^{-1}$ cm$^{-2}$]} &   \colhead{L$_x$ [erg s$^{-1}$]}  &   \colhead{L$_x$/L$_{bol}$}  &   \colhead{Ref}
    }
    \startdata
    WASP-77A    &   15709   &   9.939  &   24    &   2.08\pm0.05\times10^{-3}   &   1.49\pm0.33\times10^{-14}  &   1.97\pm0.44\times10^{28}   &   6.97\pm1.55\times10^{-6}  &   1\\
    WASP-17 &   23322   &   23.84   &   4   &   <3.78\times10^{-4}  &   <2.85\times10^{-14} &   <5.73\times10^{29}  &   <3.65\times10^{-5}  &   2\\
    LP 791-18   &   23320   &   23.79   &   2   &   <3.78\times10^{-4}  &   <2.63\times10^{-14} &   <2.13\times10^{27}  &   <2.79\times10^{-4}  &   2\\
    TOI-193 &   23321   &   22.89   &   1   &   <3.92\times10^{-4}  &   <2.60\times10^{-14} &   <1.99\times10^{28}   &   <7.68\times10^{-6} &   2\\
    \enddata
    \tablenotetext{a}{Upper limits represent the count rate required to produce a 3$\sigma$ detection given the listed exposure time}
    References: (1) \citet{salzChandraprop}, (2) \citet{meatsXrayprop}
\end{deluxetable}

Young, rapidly rotating stars show a saturation at $\log{L_x/L_{bol}}\sim-3$, and older stars begin to show a sharp decline in fractional X-ray luminosity after reaching rotation periods of a few days \citep{wright2011,astudillo2017}. Here, we take ``active'' stars to be broadly defined as those with $-5 \lesssim \log{L_x/L_{bol}} \lesssim -3$ and ``inactive'' as those with $\log{L_x/L_{bol}} < -5$ \citep[][and references therein]{linsky2019}. Under this definition we find that 9 of the 11 stars from this work exhibit low fractional X-ray luminosities consistent with low coronal activity. Note that the X-ray flare of L 98-59, indicated by the black square in figure \ref{fig:X-ray}, pushes the star into the active regime, highlighting the importance of taking into account stellar variability, as we discuss in section \ref{sec:flare}.

\subsection{FUV to NUV flux ratios}
\label{sec:FUV_NUV_ratio}
The FUV to NUV flux ratio is an important measure for exoplanet habitability studies. As described in section \ref{sec:intro_terrestrial}, the FUV/NUV ratio can impact atmospheric oxygen chemistry via the Chapman reactions. However, for a star with a large FUV/NUV flux ratio, destruction of O$_3$ via NUV flux may not balance its creation from FUV flux, leading to an abiotic buildup of ozone and the detection of a potential false-alarm biosignature \citep{segura2010,domagal2014,schwieterman2018}. 

\begin{figure}[h]
    \centering
    \includegraphics[width=0.8\textwidth]{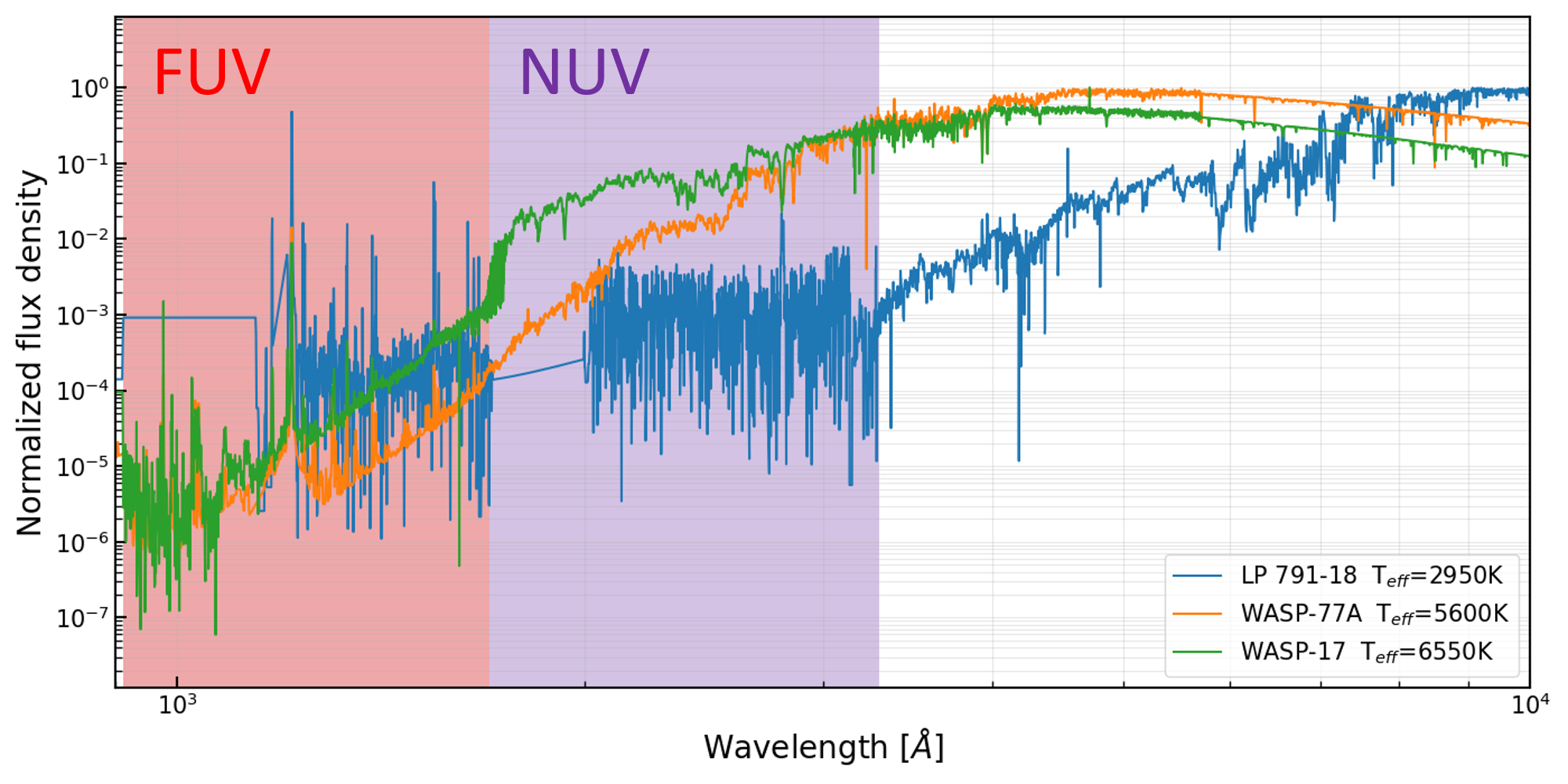}
    \caption{Stellar UV flux with increasing effective temperature. Note that increasing effective temperature drives the photospheric emission deeper into the UV as discussed in section \ref{sec:FUV_NUV_ratio}}
    \label{fig:fuv_teff}
\end{figure}

While a larger fraction of the NUV is contributed by the stellar photosphere, the FUV/NUV ratio can also be thought of as a chromospheric activity indicator, as most of the stellar FUV flux from GKM stars comes from emission lines as a result of stellar activity rather than continuum emission. It is important to keep in mind, however, that this ratio is strongly correlated with effective temperature. As $T_{\text{eff}}$ increases, the photospheric emission of the star begins to push further into the NUV region, decreasing the FUV/NUV ratio. There is a minimum in the FUV/NUV ratio at $\sim1$ M$_\sun$ after the photospheric emission begins to push all the way into the FUV, resulting in an increase of FUV/NUV flux as shown in figure \ref{fig:fuv_teff}. This makes the FUV/NUV ratio less accurate as an activity indicator for hotter stars without subtracting the photospheric contribution. However, at a given stellar mass, the FUV/NUV ratio is a measure of the excess FUV emission contributed by chromospheric and transition region activity.

\begin{figure}[h]
    \centering
    \includegraphics[width=0.8\textwidth]{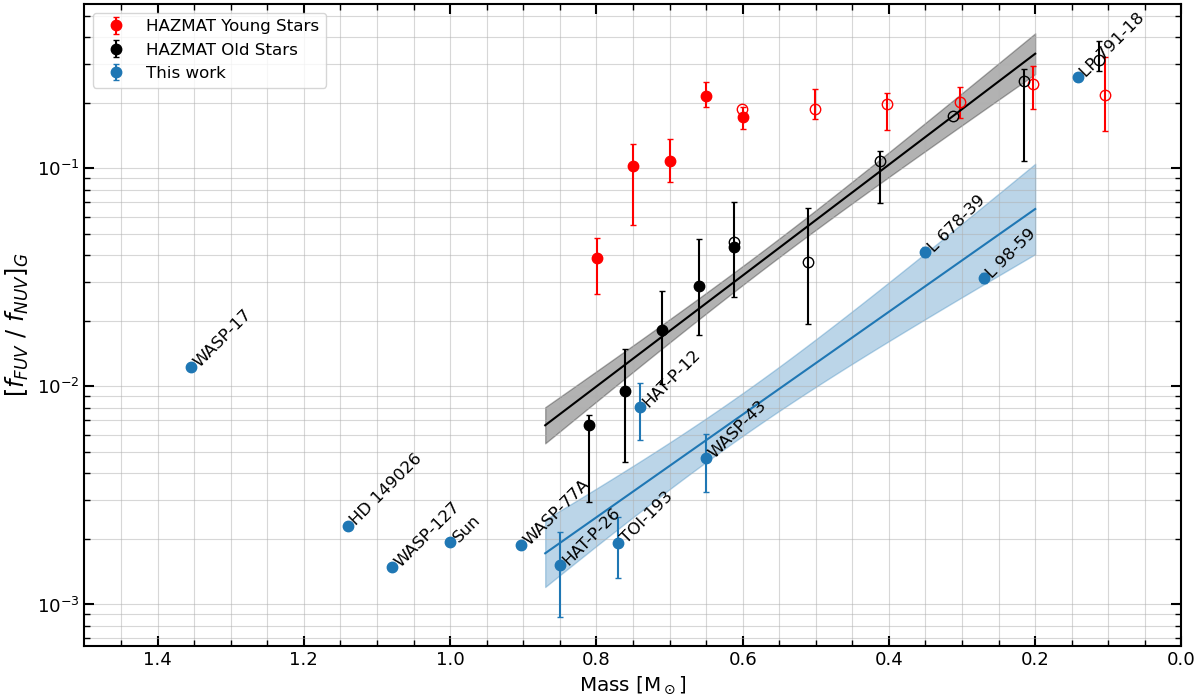}
    \caption{MUSCLES Extension stars compared with M and K dwarfs from the HAZMAT surveys. For stars from the HAZMAT surveys, open markers indicate M dwarfs from \citet{schneider2018} while solid markers represent K dwarfs from \citet{yowell2019}. Shaded regions represent the 1$\sigma$ error level of the linear fits.}
    \label{fig:HAZMAT_comp}
\end{figure}

To put the MUSCLES Extension stars in context with the general stellar population, we compare them to the HAZMAT III \citep{schneider2018} and HAZMAT V \citep{yowell2019} surveys. Comparison to the HAZMAT study offers an opportunity to compare stars selectively chosen as exoplanet-hosts (MUSCLES Extension) to a non-selective field survey (HAZMAT). The HAZMAT III and HAZMAT V surveys present studies of 642 M-dwarfs and 455 K-dwarfs, respectively. Figure \ref{fig:HAZMAT_comp} shows the FUV/NUV flux ratio in the GALEX bandpasses (FUV: 1350-1750 $\text{\AA}$, NUV: 1750-2800 $\text{\AA}$) of the MUSCLES Extension stars compared to the two HAZMAT surveys which have been grouped to 10 stellar mass bins and separated by age, with ``old" stars being those with an age of $\sim5$ Gyr \citep[see Figure 12 of][]{yowell2019}. We performed a linear fit using a least-squares routine to fit the stars between 0.2-0.85M$_{\odot}$. The bounds of the fit exclude stars in the saturated FUV/NUV regime of $\sim2\times10^{-1}$ at M $< 0.2$ M$_\sun$, as well as the MUSCLES Extension stars with spectral type of F or G for consistency with the HAZMAT surveys which include only M and K dwarfs. The slopes of the linear fits are consistent within the 1$\sigma$ level: $-2.54\pm0.2$ for HAZMAT stars and $-2.36\pm0.5$ for MUSCLES Extension stars. However, the MUSCLES Extension stars are considerably less FUV luminous than the HAZMAT stars, offset in their FUV/NUV ratio by $\sim3\sigma$, suggesting that the MUSCLES Extension targets have significantly less FUV activity than the average populations of old K and M dwarfs in the field. This can be explained by a selection bias from the techniques used to detect exoplanet systems. RV and transit surveys largely select for low-activity stars, as active stars add excess noise to RV and transit measurements which can be mistaken for planetary signals \citep{butler2017}.

\subsection{UV Ion ratios}
\label{sec:ion_ratios}
\begin{figure}[h]
    \centering
    \includegraphics[width=\textwidth]{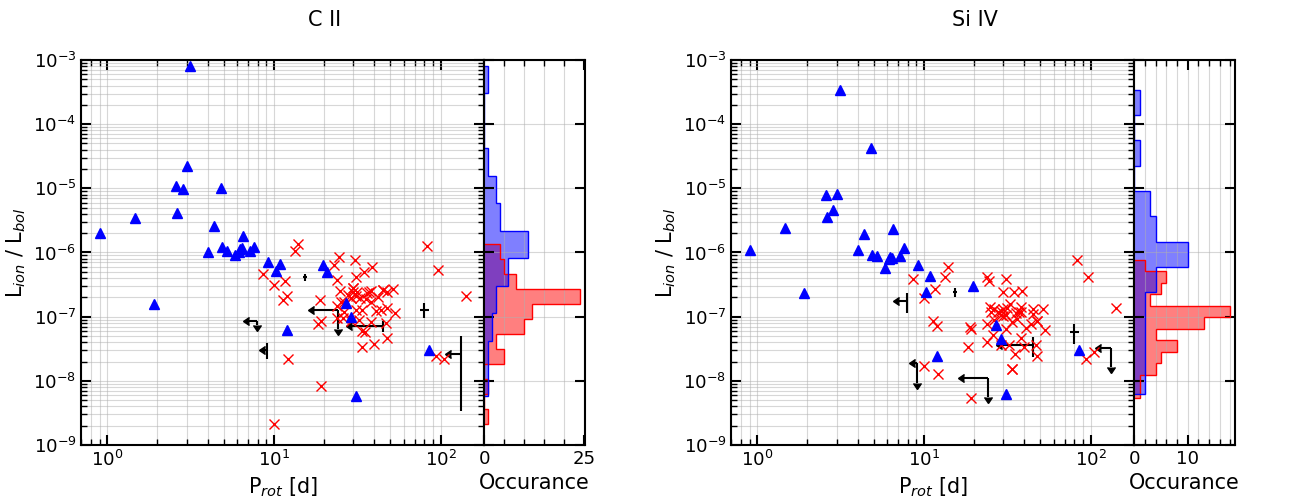}
    \\
    \includegraphics[width=\textwidth]{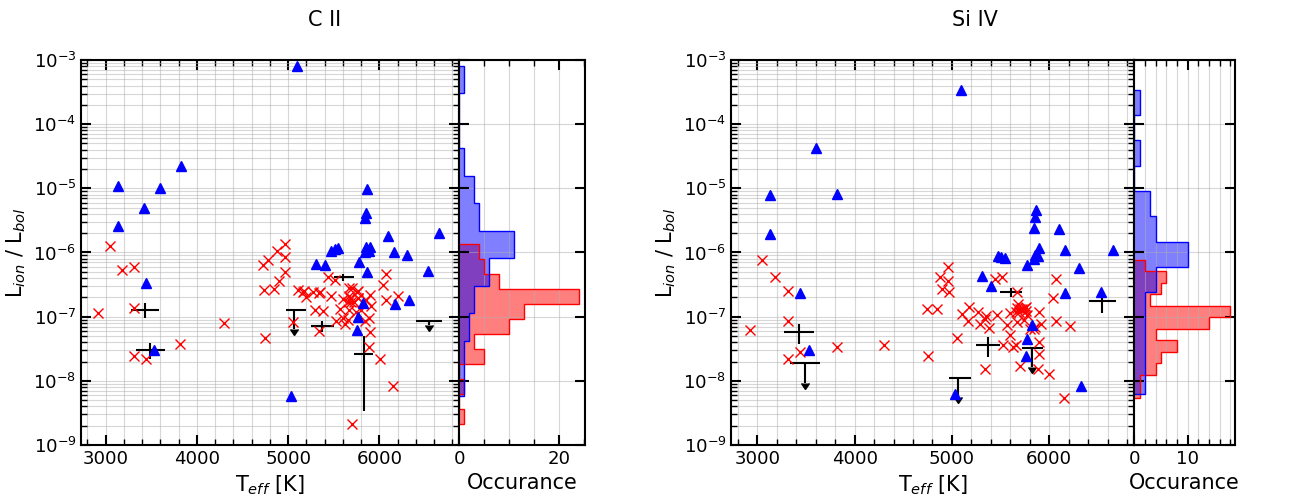}
    \caption{Fraction of bolometric luminosity of UV ions vs (a) rotation period (b) effective temperature. \textcolor{red}{$\times$} = \citet{france2018} planet hosts, $\textcolor{blue}{\blacktriangle}$ = \citet{france2018} non-planet hosts, + = this work. Histograms represent the number of stars with a bolometric luminosity ratio within the range of each bin, with blue corresponding to non-planet hosts and red to planet hosts.}
    \label{fig:ion_rats}
\end{figure}

We quantify chromospheric and transition region activity via the fraction of bolometric luminsoity, $L_{ion}/L_{bol}$, for C II ($\lambda\lambda$1334.5,1335.6 $\text{\AA}$) and Si IV ($\lambda\lambda$1393.8,1402.8 $\text{\AA}$). Figure \ref{fig:ion_rats} shows the MUSCLES Extension targets plotted against both planet and non-planet hosting stars from the \citet{france2018} survey.

The histograms of bolometric luminosity ratios for planet and non-planet hosting stars suggest two different populations. Means and standard deviations for both populations are listed in Table \ref{tab:histmeans}. We apply a two sided Kolmogorov-Smirnov (KS) test to the following groups: (1) All non-planet hosting stars and all planet hosting stars, (2) non-planet hosting stars and MUSCLES Extension stars, (3) MUSCLES Extension stars and \citet{france2018} planet hosting stars. The KS test is applied with the null hypothesis that the two sample groups come from the same parent distribution and the alternative that they have differing distributions. We find that the populations of non-planet hosts and planet hosts for both groups 1 and 2 differ at the $3\sigma$ level or greater for C II and Si IV, indicating that stars chosen from exoplanet surveys have statistically different activity levels than than the general population. We do not reject the null hypothesis that MUSCLES Extension stars come from the same distribution than the \citet{france2018} planet hosting stars, suggesting that the MUSCLES Extension targets are similar in activity level to the other planet hosting stars in our test.

\begin{deluxetable}{cCCCC}[h]
\tablecaption{Mean and standard deviations of C II and Si IV fraction of bolometric luminosity\label{tab:histmeans}}
\tablehead{\colhead{Stellar population}    &   \colhead{C II mean}  &   \colhead{C II standard deviation}   &   \colhead{Si IV mean}    &   \colhead{Si IV standard deviation}}
\startdata
Non-planet hosts    &   2.75\times10^{-5}   &   1.40\times10^{-4}   &   1.39\times10^{-5}   &   6.09\times10^{-5}\\
Planet hosts    &   2.47\times10^{-7}   &   2.64\times10^{-7}   &   1.34\times10^{-7}   &   1.40\times10^{-7}\\
\enddata
\end{deluxetable}


\subsection{UV irradiation environment of exoplanets}
\label{sec:planet_flux}
\begin{figure}[h]
    \centering
    \includegraphics[width=0.95\textwidth]{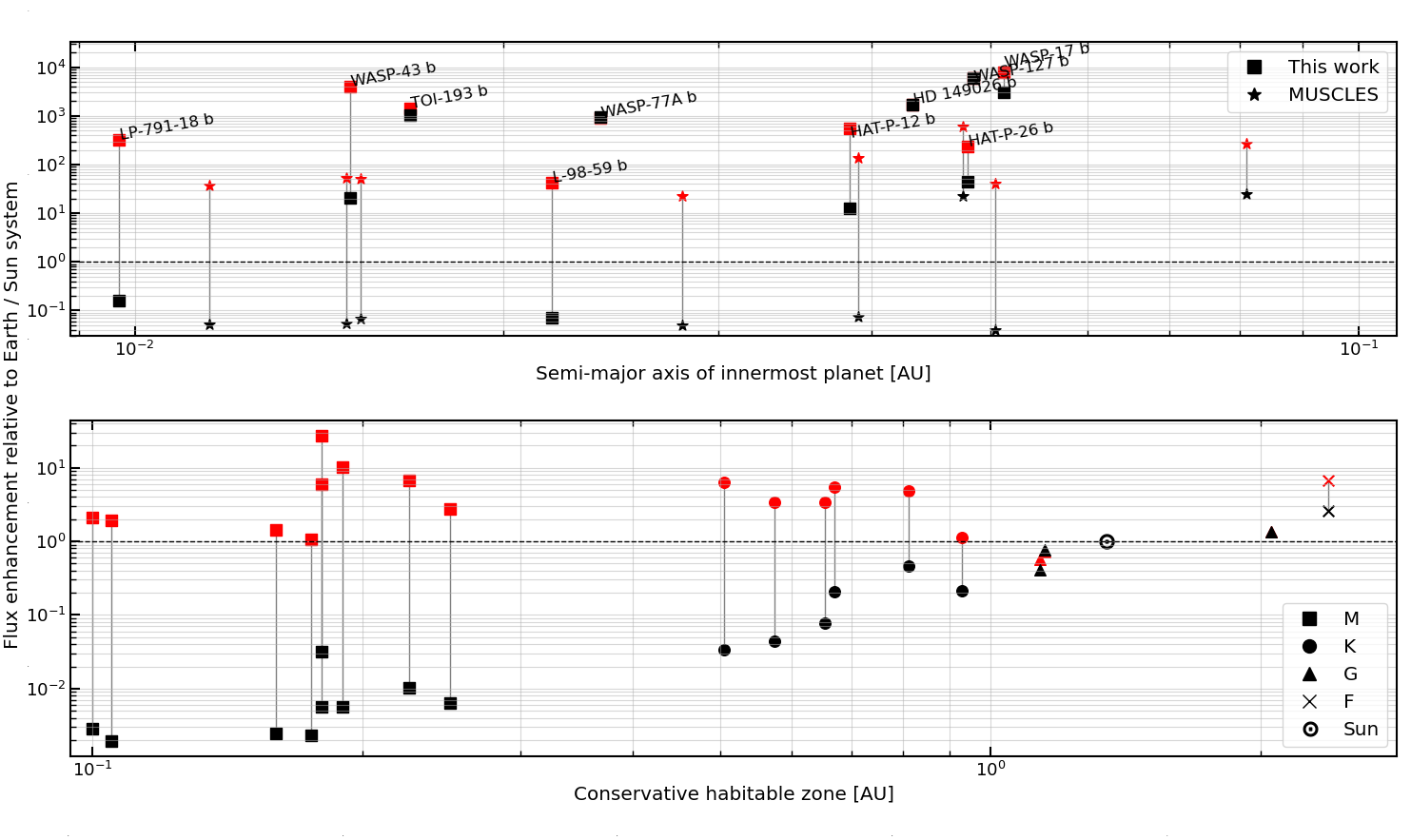}
    \caption{(Top) UV flux incident at the top of the innermost planet's atmosphere relative to the UV flux incident at Earth. (Bottom) Amount of flux incident at the conservative habitable zone (average of the inner- and outermost habitable zones using \citet{kopparapu_HZ} models) relative to the flux experienced at the Sun's habitable zone of approximately 1.3 AU. For both panels red markers indicate flux over the FUV band and black markers the NUV band.}
    \label{fig:planet_flux}
\end{figure}

Here we present our FUV and NUV flux measurements in context of the closest orbiting planet for each system. Figure \ref{fig:planet_flux} shows the incident top-of-the-atmosphere FUV and NUV flux relative to the Earth/Sun system for the MUSCLES Extension targets and a subset of the original MUSCLES targets as well as flux incident at each stellar system's conservative habitable zone, calculated as the average of the inner and outer habitable zone using the models of \citet{kopparapu_HZ}. Stars with a solar-like FUV/NUV ratio are those with little separation on the plot, while those with FUV/NUV ratios much different than the Sun (e.g., M dwarfs) have a large separation between FUV and NUV.

Many previous studies have used M dwarfs observed by the MUSCLES program as inputs to atmospheric models of hypothetical planets (e.g., \citet{tian2014,rugheimer2015,chen2021}). The MUSCLES Extension is the first time that FUV and NUV fluxes have been directly measured for an ensemble of stars whose planets have been or will be characterized with high-sensitivity spectrophometric observations, allowing for more accurate modeling of photochemical contributions to atmospheric composition and evolution.

Planets around our G and F stars experience flux enhancements of $\sim10^3-10^4$ relative to Earth/Sun in both the FUV and NUV, due to their smaller semi-major axes. Notably, the smaller M and K stars show FUV enhancements of $\sim1-2$ orders of magnitude more that of the Earth/Sun but NUV decrements of $\sim1$ order of magnitude \textit{less} than the Earth/Sun system. This indicates a strong possibility of photochemical disequilibrium, as photodissociation rates in the FUV will be enhanced while those in the NUV, including O$_3$, will be suppressed relative to Earth.

\subsection{L 98-59 X-ray flare}
\label{sec:flare}
\begin{figure}[h]
    \centering
    \includegraphics[width=0.8\textwidth]{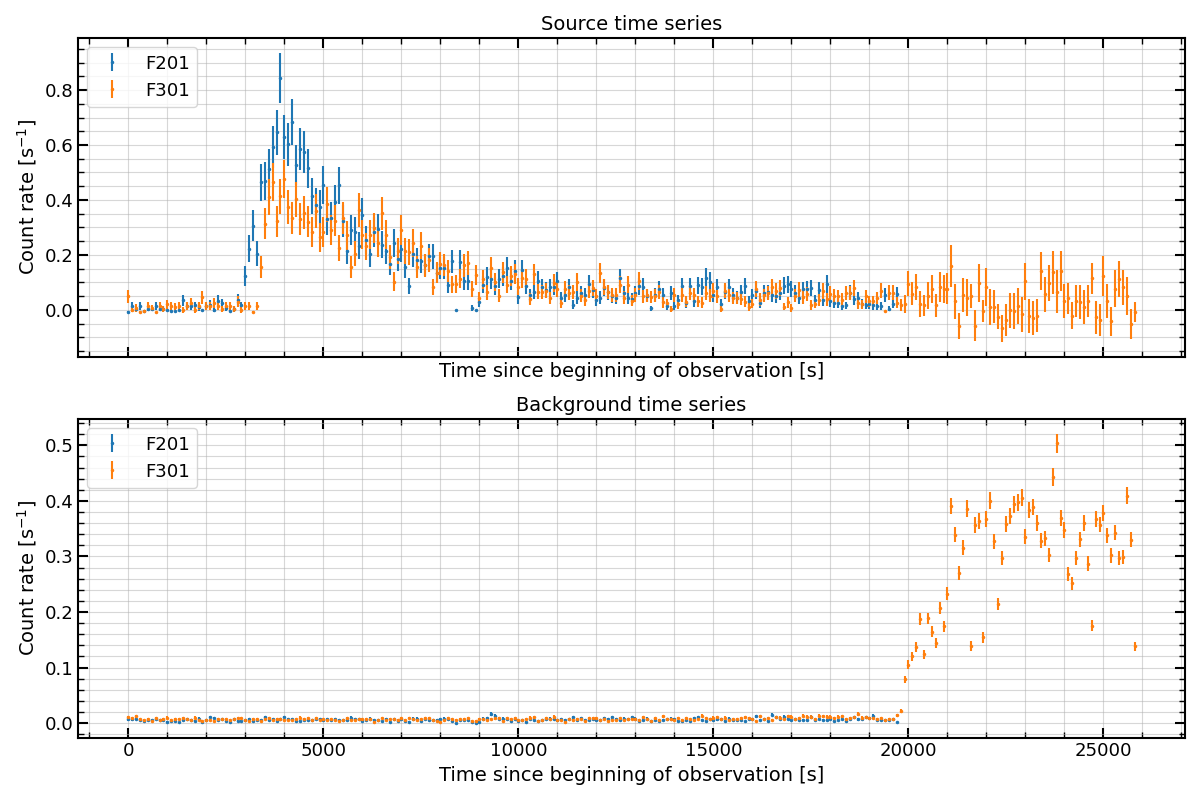}
    \caption{Top: Background subtracted source light curves for F201 and F301. Bottom: Background time series for each observation. Note the large background flare occurring at 20 ks in F301; this is likely due to protons within the Earth's magnetosphere being funneled towards the detector. All data are from the EPIC pn detector with the medium filter and binned to 100s.}
    \label{fig:light_curve}
\end{figure}

We obtained two \textit{XMM-Newton} observations of L 98-59 which each had an X-ray flaring event. The first was on April 15, 2021, 08:26:45 UTC (obs.ID 0871800201, hereafter F201) and the second on April 16, 2021, 03:27:28 UTC (obs.ID 0871800301, hereafter F301). Both flares were detected in the EPIC pn and MOS detectors (X-ray; $\sim0.3-10.0$ keV) and F301 was also detected in the Optical Monitor with UVW1 filter (OM UVW1; 2000-4000 $\text{\AA}$). We extracted the light curves using the SAS \textit{evselect} routine with a temporal bin size of 100s. The two flares are shown together, relative to the beginning of their respective observations, in Figure \ref{fig:light_curve}. Defining the start time as the point where the light curve first exceeds the quiescent level, and the rise time as the difference between the time of peak count rate and the start time, we find that F301 had a start time of $3300\pm50\text{ s}$ and a rise time of $400\pm70\text{ s}$. F201 had a start time of $2900\pm50\text{ s}$ and a longer rise time of $1000\pm70\text{ s}$.

L 98-59 does not return to its quiescent level during the length of the observation for either flare; thus, we are unable to quantify a definitive duration. We report instead a lower limit duration which is the time from flare start to the end of the usable observation. The flare durations are $>15.7$ ks for F301 and $>16.8$ ks for F201.

\subsubsection{Flare luminosity and equivalent duration}
\label{sec:energy_eqdur}
The X-ray flare of L 98-59 produced enough counts to allow for spectral fitting. We use the XSPEC software \citep{xspec} with an VAPEC plasma model to model the stellar spectrum. Due to the time variable plasma properties during a flaring event, spectral analysis of the flares was conducted by splitting each flare into four time intervals---quiescent, peak, decay, and tail regions. ``Quiescent" is the pre-flare region, ``peak" the region of $\sim3$ ks centered on the peak of the flare, ``decay" the region of declining count rate until the rate begins to reach a constant value at $\sim10$ ks, and ``tail" the remainder of the exposure. We use a single or double temperature VAPEC model to derive a characteristic plasma temperature for each time interval with metal abundances as described in the study of the M dwarf system LTT 1445 by \citet{brown2022}. The interstellar hydrogen column density was fixed at $1\times10^{18}$ cm$^{-2}$. The resulting flux from the best fit VAPEC models were used to calculate the luminosities of each flare as

\begin{equation}
    L_x=4\pi d^2\int{FdE}
\end{equation}

\noindent where $0.3 < E < 10.0$ keV, $d$ is the stellar distance, and $F$ the flux from the best-fit VAPEC models. The quiescent luminosities from the best-fit VAPEC models were $L_x=3.82\pm0.70\times10^{26}$ erg s$^{-1}$ and $L_x=4.13\pm0.57\times10^{26}$ erg s$^{-1}$ for F201 and F301, respectively. The calculated flare luminosities were $L_x=2.05\pm0.05\times10^{28}$ erg s$^{-1}$ for F201 and $L_x=1.59\pm0.40\times10^{28}$ erg s$^{-1}$ for F301. In comparison, a recent study of Proxima Centauri by \citet{fuhrmeister2022proxcen} finds that during an average flare the ratio of peak count rate to quiescent count rate is 10 and the average flare luminosity is $L_x=6.7\times10^{27}$ ergs s$^{-1}$. The VAPEC parameters and X-ray properties of both observations are listed in table \ref{tab:L98}.

\begin{deluxetable}{cCCCC}[h]
\tablecaption{Properties of L 98-59 X-ray flares\label{tab:L98}}
\tablehead{
\colhead{}  &   \colhead{0871800201 (F201) - Flare}    &   \colhead{0871800301 (F301) - Flare}    &\colhead{F201 - Quiescent}   &   \colhead{F301 - Quiescent}}
\startdata
Exposure time [ks]   &   14.039 &   13.819  &   2.896   &   2.896\\
Net counts    &   2298  &   1838   &   36  &   49\\
Count rate [cnt s$^{-1}$]\tablenotemark{\scriptsize a}    &   0.502\pm0.002 &   0.295\pm0.001   &   0.010\pm0.002   &   0.012\pm0.002\\
f$_x$ [10$^{-13}$ erg s$^{-1}$ cm$^{-2}$]  &   17.1\pm0.4 &   13.2\pm0.3    &   0.32\pm0.06   &   0.35\pm0.05\\
$\log{L_x}$ [erg s$^{-1}$]   &   28.3\pm0.01 &   28.2\pm0.01    &   26.6\pm0.08    &   26.6\pm0.06\\
$\L_x/L_{bol}$ &   4.68\pm0.02\times10^{-4}    &   3.63\pm0.02\times10^{-4}   &   8.72\pm0.17\times10^{-6}   &   9.43\pm0.13\times10^{-6}\\
$L_x$ enhancement\tablenotemark{\scriptsize a}   &   53.7    &   38.5    &   \text{---}  &   \text{---}\\
kT$_1$ [keV]\tablenotemark{\scriptsize b}  &   1.21^{+0.157}_{-0.169}   &  1.32^{+0.169}_{-0.189}   &   \text{---}  &   \text{---}\\
kT$_2$ [keV]\tablenotemark{\scriptsize b}   &   0.268^{+0.095}_{-0.047} &   0.294^{+0.128}_{-0.055} &   \text{---}  &   \text{---}\\
$[\text{Fe/H}]$\tablenotemark{\scriptsize b} &   0.215^{+0.147}_{-0.167} &   0.302^{+0.347}_{-0.202} &   \text{---}  &   \text{---}\\
\enddata
\tablenotetext{a}{Enhancement represents the ratio of X-ray luminosity during the peak region to the quiescent luminosity}
\tablenotetext{b}{Reported value is during the flare peak}
\end{deluxetable}

We also compute the equivalent duration, $\delta$, of each flare. The equivalent duration represents the amount of time it would take the star, in the quiescent state, to release the same amount of energy as is released during the flare \citep{gershberg1972,huntwalker2012}:

\begin{equation}
    \label{eq:equivalent_duration}
    \delta = \int_{t_0}^{t_0+\Delta t}\frac{(R_f-R_q)}{R_q}dt
\end{equation}

\noindent where $t_0$ is the start time of the flare, $\Delta t$ the flare duration, $R_f$ the 0.3-10 keV count rate during the flare, and $R_q$ the 0.3-10 keV count rate during the quiescent period. The equivalent durations of the L 98-59 flares were $\delta>130.4$ ks for F301 and $\delta>245.9$ ks for F201. These are larger than the equivalent durations reported in \citet{loyd2018} which had values $1.3 < \delta < 120.9$ ks for similarly inactive M and K stars. However, 2 of 3 flares from the MUSCLES study were also truncated before the end of the flare duration and thus are also considered to be lower limits.

Accounting for the flares increases $L_x/L_{bol}$ to 38.5 and 53.7$\times$ the quiescent level. This highlights the importance of accounting for high-energy flaring events in the XUV radiation environment of exoplanets around M dwarfs, especially as we have observed two similar events within a 24 hr period.

\subsubsection{NUV flare of F301}
\label{sec:U-band}

The OM UVW1 flare from F301 peaks during the rise time of the X-ray flare, shown in figure \ref{fig:neupert}. {\citet{gudel2002} report a similar phenomenon during a flaring event of Proxima Centauri, for which they also obtained simultaneous \textit{XMM-Newton} EPIC and OM UVW1 observations. \citet{gudel2002} propose a chromospheric evaporation scenario similar to the well-studied Neupert effect \citep{neupert1968} in which a flaring event accelerates high-energy electrons into the chromosphere, where the subsequent deposition of energy causes a sharp optical signal and increase in chromospheric temperature driving hot plasma into the corona and resulting in the longer duration X-ray flare. The Neupert effect was first described in relation to simultaneous hard X-ray and microwave flares in the solar corona but had not been seen in the X-ray and NUV/optical prior to its observations on other stars. The X-ray/optical relation has since been observed in several stellar spectra of M and K dwarfs: in BY Draconis (K4) \citep{dejager1986}, UV Ceti (M6) \citep{dejager1986,schmitt1993}, AD Leonis (M3) \citep{hawley1995}, Proxima Centauri (M5.5) \citep{gudel2002,gudel2004}, and now L 98-59 (an ``optically inactive" M3). If the chromospheric evaporation scenario proposed by \citet{gudel2002} is correct, one should expect the time derivative of the X-ray light curve to mimic the optical signal:

\begin{equation}
    \label{eq:neupert}
    \frac{dL_x}{dt}\propto L_O
\end{equation}

Figure \ref{fig:neupert} shows light curves from the EPIC and OM UVW1 instruments during the F301 flaring event. We are not concerned with total counts in this analysis and as such we have rebinned both X-ray and NUV light curves to 10s intervals rather than 100s to obtain more accurate temporal measurements. As expected from Equation (\ref{eq:neupert}), the time derivative of the X-ray lightcurve matches the shape of the OM UVW1 light curve (Figure \ref{fig:neupert}b), with a difference in peak timing of 27$\pm7$s. This supports the theory of chromospheric evaporation occuring during stellar flaring events.

\begin{figure}[h]
    \centering
    \includegraphics[width=0.8\textwidth]{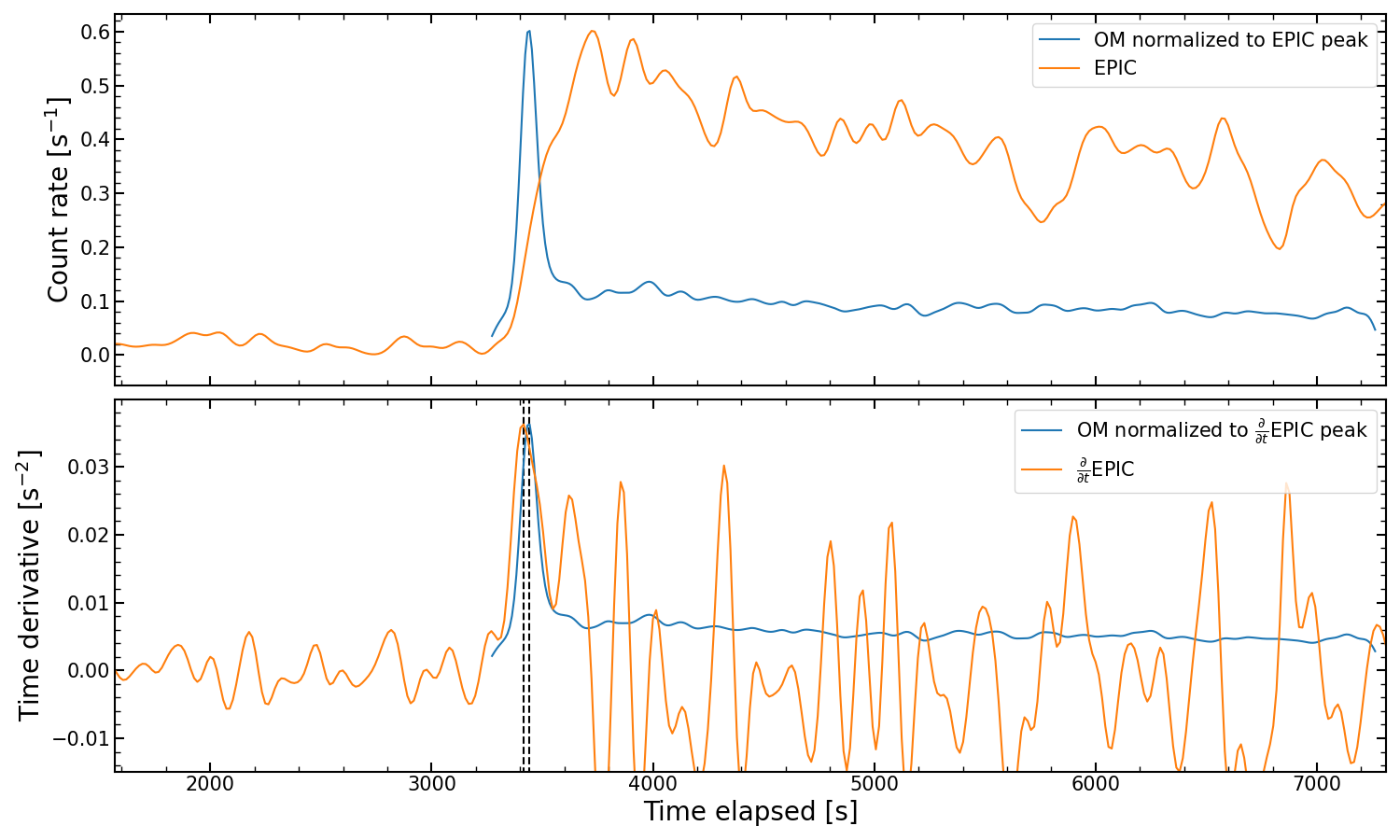}
    \caption{L 98-59 X-ray and NUV light curves for F301. The EPIC light curves have been binned to 10s to match the time resolution of the OM UVW1 data. The data have been smoothed with a Gaussian kernel for clarity. Top: OM UVW1 flare normalized to the peak of the X-ray flare. Bottom: Time derivative of the X-ray flare (units of s$^{-2}$) overplotted by the normalized OM UVW1 light curve (units of s$^{-1}$). Note that the blue OM UVW1 curve does not match the units of the vertical axis and is only plotted here as a visual aid. The vertical black lines represent the location of the peak of each light curve.}
    \label{fig:neupert}
\end{figure}

\section{Summary}
\label{sec:summary}
The MUSCLES Extension for Atmospheric Transmission Spectroscopy is a study of UV and X-ray activity of 11 exoplanet hosting stars whose systems are to be observed as part of the JWST ERS and GTO programs but which have no previous UV characterization. We obtained FUV and NUV observations of each target using \textit{HST}-STIS and -COS, as well as a combination of new and archival X-ray observations using \textit{XMM-Newton} and \textit{Chandra} observatories. We assessed the chromospheric and transition region activity levels of each star based on their FUV/NUV flux ratios and fraction of bolometric flux from X-rays and UV ion emission lines and put these activity measurements into context by comparing the MUSCLES extension targets to a broader population of planet and non-planet hosting stars.

The main results of this work are:
\begin{enumerate}
    \item We have assembled panchromatic SEDs from 5.5 $\text{\AA}$-5 $\mu$m of 11 exoplanet hosting stars with guaranteed \textit{JWST} observation time but no prior UV characterization with \textit{HST}. The SEDs will be available as high-level science products on the MUSCLES portal hosted on the MAST archive\footnote{\url{https://archive.stsci.edu/prepds/muscles/}} and can be used as inputs for the stellar irradiance when modeling planetary atmospheres observed by \textit{JWST}, eliminating the need to rely on optical scaling relations or stellar models without a complete treatment of the upper stellar atmospheres.
    \item The planet-hosting stars from our survey follow the trend of \citet{france2018}, displaying statistically lower activity levels than non-planet hosting groups on the basis of fractional X-ray luminosity, UV ion emission, and FUV/NUV flux ratios. This can easily be explained by a sample bias: confirmed exoplanets from RV and transit surveys largely select for low-activity stars, as active stars add excess noise to RV and transit measurements. However, as planet population estimates expect to find exoplanets around the vast majority of stars, selecting stellar irradiance levels based on samples of known exoplanet-host stars likely underestimates the UV flux experienced over a planet's lifetime and is not indicative of the radiation environments of the exoplanet population at large. Therefore, we present a cautionary speculation that the UV-driven atmospheric photochemistry on the average galactic exoplanet may be significantly different than what we measure on the current set of planets being studied by \textit{JWST}.
    \item We find two X-ray flares on the M dwarf L 98-59 which increased the X-ray fraction of its bolometric luminosity by factors of 38.5 and 53.7 times the quiescent levels. Depending on the frequency of flaring events, this could significantly increase the total amount of XUV irradiation of the planets orbiting this otherwise inactive star. This highlights the importance of studying time variability in exoplanet-hosting stars in order to accurately model a planet's lifetime-integrated UV irradiance.
\end{enumerate}

\begin{acknowledgments}
The HST observations presented here were acquired as part of the Cycle 28 program 16166, supported by NASA through a grant from the Space Telescope Science Institute, which is operated by the Association of Universities for Research in Astronomy, Inc., under NASA contract NAS 5–26555. This work is based in part on observations made by the Chandra X-ray Observatory, supported by Chandra grant G01-22022X to the University of Colorado, and by data obtained from the Chandra Data Archive, and based on observations obtained with XMM-Newton, an ESA science mission with instruments and contributions directly funded by ESA Member States and NASA. 

This research has made use of the NASA Exoplanet Archive, which is operated by the California Institute of Technology, under contract with the National Aeronautics and Space Administration under the Exoplanet Exploration Program.
\end{acknowledgments}

%

\vspace{5mm}
\facilities{HST, XMM-Newton, Chandra}


\software{astropy \citep{2013A&A...558A..33A,2018AJ....156..123A},  
          scipy \citep{scipy}, 
          matplotlib \citep{matplotlib},
          SAOImage DS9 \citep{ds9},
          Scientific Analysis System \citep{SAS},
          CIAO \citep{CIAO},
          XSPEC \citep{xspec}
}


\appendix
\section{Targets}
\label{sec:targets}
The targets for the MUSCLES Extension include 3 M, 3 K, 5 G, and 1 F type dwarf stars. Notable differences from the original MUSCLES and Mega-MUSCLES survey are the inclusion of G and F type stars and the much larger distances to the targets. Our targets host a wide variety of exoplanets, from sub-Earth size through giant hot-Jupiters. This section is dedicated to providing a brief description of each planetary system as well as the spectral data used during construction of the SEDs.

\subsection{WASP-17}
WASP-17 is an F6 star at a distance of approximately 405 pc based on Gaia Data Release 3 \citep[DR3;][]{gaiaDR3} data. The star is estimated to have a rotation period of 8.5-11 days and an age of approximately 2.7 Gyr with a sub-Solar metallicity of $\text{[Fe/H]}=-0.190\pm0.090$ \citep{bonomo2017}. The system consists of one confirmed and peculiar exoplanet. WASP-17b is an ultra-low density planet with a radius of R$_{\text{p}}\approx 2$R$_{\text{J}}$ but a mass of only M$_{\text{p}}\approx0.5$M$_{J}$. Initial observations of WASP-17b \citep{anderson2010} suggested the planet has a retrograde orbit; this was later confirmed by \citep{bayliss2010}. The proposed explanation for WASP-17b's retrograde orbit is a combination of planet-planet scattering, the Kozai mechanism, and tidal circularization \citep[][and references therein]{anderson2010}.

We obtained STIS G140L, G230L, and G430L observations of WASP-17. The G140L spectra showed no evidence of FUV emission line flux and had poor quality over the Lyman-$\alpha$ region. We did not obtain G140M spectra and therefore were unable to reconstruct the Lyman-$\alpha$ emission line using MCMC methods but the strong S/N in the NUV G230L observations allowed for estimation using the Mg II relation.

We obtain 23.84 ks of new \textit{Chandra} ACIS-S observations (obs.ID 201354, PI France) of WASP-17. Our X-ray analysis of the target was a non-detection and we present a 3$\sigma$ upper limit for our further X-ray analysis as discussed in section \ref{sec:xray}.

The proxy star used for the X-ray and FUV spectrum was Procyon, an F5 star with $T_{\text{eff}}\approx7740$ K. We obtained a high resolution UV spectrum of Procyon from the SISTINE II sounding rocket observation (Cruz-Aguirre et al. - in prep). The effective temperature of Procyon is $\sim1200$ K higher than WASP-17 but we find that the NUV scaling procedure matched the shape of the spectrum very well in the G230L region from 1750-3150 $\text{\AA}$ and we find that Procyon has $L_x/L_{bol}\sim1\times10^{-5}$, similar to the upper limit of WASP-17. Thus, we believe that Procyon is a suitable proxy star for WASP-17.

\subsection{HD 149026}
HD-149026 is a G0 star with a DR3 distance of approximately 76 pc. The star has a super-Solar metallicity, [Fe/H]=$0.36\pm0.05$, and a single confirmed exoplanet with an unusually dense core \citep{sato2005}. The exoplanet, HD 149026b, has a radius of $R=0.725\pm0.05$ $R_{J}$ but a density of 1.7 times that of Saturn. The high metallicity of the system in conjunction with the large density of the planet indicate that it may have an icy/rocky core that makes up 50-80\% of the planetary mass \citep{sato2005,fortney2006}.

We obtained STIS G140L, G140M, E230M, and G430L observations of HD-149026. During the G140L observations, the shutter door remained closed for the entire duration and thus no spectra were acquired. Of the two G140M observations, the second had a misplaced extraction box during the X1DCORR routine and had to be re-extracted. After re-extraction we found the S/N of the G140M observations to be sufficient to reconstruct the Lyman-$\alpha$ emission line. The E230M spectrum has S/N above the threshold for all wavelengths and has been convolved to match the resolution of the G230L spectra used in the rest of the SEDs. The G430L observation has S/N above the threshold for all wavelengths $\lambda\gtrsim3050$ $\text{\AA}$.

We retrieved 10.8 ks of archival \textit{XMM-Newton} observations (obs.ID 0763460301, PI Salz) in which the target is detected and we find an x-ray luminosity of $\log{L_x}=27.6\pm0.1\times10^{26}$ erg s$^{-1}$.

Due to similar spectral type and rotation period, the proxy star for this target is the quiet Sun \citep{woods2009}.

\subsection{WASP-127}
WASP-127 is a G5 star with an estimated age of approximately 11 Gyr \citep{lam2017} and DR3 distance of 160 pc. Its planetary companion, WASP-127b, has an anomalously low density, with a sup-Saturn mass of $M=0.18\pm0.02$ $M_J$, super-Jupiter radius of $R=1.37\pm0.04$ $R_J$, and orbital period of 4.17 days, WASP-127b falls within the previously discussed sub-Neptune desert \citep{lam2017,skaf2020}. Transmission spectroscopy of WASP-127b shows a feature-rich spectrum including absorption by Na, H$_2$O, and either CO$_2$ or CO \citet{spake2021}. Additionally, the low density ``puffiness'' of WASP-127b's atmosphere is unlikely to be caused by photo-evaporation due to its host star's low UV flux \citep{palle2017,skaf2020} making it an interesting target for alternative atmospheric inflation processes \citep[see][and references therein]{skaf2020}.

We obtained STIS G140L, G140M, G230L, and G430L observations of WASP-127. The FUV G140L spectrum showed no evidence of FUV emission lines and the Lyman-$\alpha$ emission from G140M was insufficient to recreate the intrinsic emission from MCMC methods so we opted to use the Mg II scaling relation. The NUV G230L spectrum breaks the S/N$>3$ threshold for wavelengths of $\lambda>2050$ $\text{\AA}$ and extremely faint Mg II emission.

We retrieved 8 ks of archival \textit{XMM-Newton} observations (obs.ID 0853380601, PI Schartel) which yielded a non-detection of the target.

The X-ray and FUV proxy for WASP-127 is the quiet Sun. Despite its higher T$_{\text{eff}}$, we chose the solar spectrum based on its similarly low chromopsheric and coronal activity levels and find that it provides a good fit to the stellar continuum below $\sim2600$ $\text{\AA}$.

\subsection{TOI-193}
TOI-193, also designated LTT 9779, is a solar-like G7 star with DR3 a distance of 81 pc. It has an estimated age of 2 Gyr and a super-solar metallicity of $\text{[Fe/H]}=0.25\pm0.04$ \citep{jenkins2020}. \citet{jenkins2020} confirmed an exoplanet, TOI-193b, with a mass of M$=9.225^{+0.25}_{-0.26}\times10^{-2}$ M$_{\text{J}}$, radius of R=$0.421\pm0.02$ R$_{\text{J}}$, and orbital period of 0.79d. Like WASP-127b, this places TOI-193b firmly within the Neptune desert, offering another opportunity to study the region between hot-Jupiters and super-Earths.

We obtained STIS G140L, G140M, G230L, and G430L observations of TOI-193. The G140L FUV observations did not show any emission lines above the S/N$>3$ threshold. We detect Lyman-$\alpha$ emission in the G140M spectrum albeit with low S/N. Despite the quality we were able to reconstruct the Lyman-$\alpha$ emission line from G140M observations. The G230L NUV spectrum breaks the S/N threshold for wavelengths of $\lambda>2200$ $\text{\AA}$ and shows faint Mg II emission within the photospheric absorption band.

We also obtained a new 22.89 ks \textit{Chandra} ACIS-S observation of TOI-193 which showed a non-detection.

The X-ray and FUV proxy star chosen was again the quiet solar spectrum based on similar UV activity level indicators.

\subsection{WASP-77A}
WASP-77A is a G8 star with a K-dwarf companion. It has a DR3 distance of 105 pc. \citet{maxted2013} report WASP-77A to be solar-like in mass, radius, and metallicity; follow-up observations by \citet{cortes2020} yield a slightly sub-solar metallicity of $\text{[Fe/H]}=-0.10^{+0.10}_{-0.11}$. \citet{maxted2013} report an age of $\sim1$ Gyr using a rotation period relation or an age of $\sim8$ Gyr using stellar models. Following studies by \citet{cortes2020,bonomo2017} report an age of $\sim6$ Gyr and a $\log{R'_{HK}}=-4.57\pm0.02$ \citep{salz2015}, indicating a low activity, sub-Solar star. The single confirmed exoplanet, WASP-77Ab, is a typical hot-Jupiter with $M=1.76\pm0.06$ $M_J$, $R=1.21\pm0.02$ $R_J$, and period of 1.36 d \citep{maxted2013}. It is estimated to have a high mass-loss rate from previous X-ray studies \citep{salz2015} and provides a promising opportunity to study hot-Jupiter planets around solar-like stars.

We obtained STIS G140L, G140M, G230L, and G430L observations of WASP-77A. We find that the STIS FUV observations have S/N $>3$ for most emission lines. However, despite good S/N in the emission lines, we were unable to reconstruct the Lyman-$\alpha$ line using the MCMC method and so report the estimated flux based on the Mg II relation. The G230L NUV observations break the S/N$>3$ threshold for wavelengths of $\lambda>2000$ $\text{\AA}$ and show Mg II in emission within the photospheric absorption band.

We retrieved 9.94 ks of archival \textit{Chandra} ACIS-S observations (obs.ID 15709, PI Salz) in which the target is detected with a fractional X-ray luminosity of $\log{L_x/L_{bol}}=-4.25$.

The X-ray and FUV proxy star was again the quiet solar spectrum based on similar X-ray and UV activity level indicators.

\subsection{HAT-P-26}
HAT-P-26 is a K1 star with a DR3 distance of 142 pc. Initial observations \citep{hartman2011} reported the star to be slightly smaller than the Sun with a similar metallicity of $\text{[Fe/H]}=-0.04\pm0.08$. The system has an age of $9.0^{+3.0}_{-4.9}$ Gyr and $\log{R'_{HK}}=-4.992$ \citep{hartman2011}; this indicates that HAT-P-26 is an old and inactive star. The exoplanet, HAT-P-26b, is a Neptune-sized planet with $M=0.059\pm0.007$ $M_J$, $R=0.565^{+0.072}_{-0.032}$ $R_J$, and period of 4.23 d \citep{hartman2011}, making it the third star from this study which falls in the Neptune desert. It is notable for its low density which is consistent with an irradiated planet with 10 $M_\earth$ rocky core and 8 $M_\earth$ gas envelope \citep[][based on \citet{fortney2007}]{hartman2011}. \citet{hartman2011} suggest that HAT-P-26b may have started its life as a Jupiter sized planet and lost $\sim30\%$ of its initial mass based on the energy-limited escape described by \citet{erkaev2007} and \citet{yelle2008}. However, \citet{hartman2011} note that due to the lack of knowledge of the XUV flux of its host star, the exact value of HAT-P-26b's mass loss is poorly constrained.

We obtained STIS observations with the G140L, G140M, G230L, and G430L gratings. Other than Lyman-$\alpha$, we find no UV emission lines with flux greater than the noise level in either the G140L or G140M observations. We also find no evidence of Mg II emission despite having good S/N beyond 2550 $\text{\AA}$ in the G230L observation. This is consistent with \citet{hartman2011}'s claim of HAT-P-26 being an inactive star. We were unable to reconstruct the Lyman-$\alpha$ emission profile and instead report an upper limit Lyman-$\alpha$ flux based on the RMS value of the continuum subtracted region over the Mg II line which we consider to be an upper limit of the Mg II flux.

We retrieved 17 ks of archival \textit{XMM-Newton} observations (obs.ID 0804790101, PI Sanz-Forcada). Our analysis of the observation showed a non-detection.

The X-ray and FUV proxy for HAT-P-26 is HD 40307, a K2.5 dwarf observed during the MUSCLES survey.

\subsection{HAT-P-12}
HAT-P-12 is a K4 dwarf with a DR3 distance of 143 pc and a sub-solar metallicity of $\text{[Fe/H]}=-0.29\pm0.05$. The single confirmed exoplanet, HAT-P-12b, first reported by \citep{hartman2009}, is a low density gas giant with mass $M_p=0.211\pm0.012$ $M_J$ and radius $R_p=0.959^{+0.029}_{-0.021}$ $R_J$, with an orbital period of 3.21 d. HAT-P-12b is found to be consistent with models of an irradiated planet with a $\lesssim10$ $M_\earth$ rocky core and a H/He dominated gas envelope \citep[][and references therein]{hartman2009}.

Due to its large distance and expected low activity, G140L and G140M exposure times required to obtain S/N greater than the threshold were prohibitively long and thus we obtained no G140L or G140M observations of HAT-P-12. Therefore, we cannot report any FUV emission line fluxes. However, we obtained both G230L and G430L observations with good S/N for wavelengths $\lambda>2500$ $\text{\AA}$ and were able to recreate the Lyman-$\alpha$ emission line based on the Mg II relation. 

We retrieved 10 ks of archival \textit{XMM-Newton} observations (obs.ID 0853380901, PI Schartel). Our analysis of the observation showed a non-detection. 

The X-ray and FUV proxy for HAT-P-12 is HD 85512, a K6 dwarf observed in the MUSCLES survey.

\subsection{WASP-43}
WASP-43 is a K7 star with a DR3 distance of 87 pc. The exoplanet, WASP-43b, was first reported by \citet{hellier2011} as a hot-Jupiter with mass and radius $M_p=2.0\pm0.1$ $M_J$ and $R_p=1.06\pm0.05$ $R_J$, respectively, orbiting very close to the host star with a semi-major axis of 0.014 AU and period of 0.813 d. Follow-up observations using TRAPPIST by \citet{gillon2012} confirmed these parameters with higher precision. Based on stellar rotation period, WASP-43 is estimated to be a young star around 0.4 Gyr \citep{hellier2011}; however, using the \citet{fortney2010} relation between radius and age for a low-irradiation planet, \citet{gillon2012} claim that WASP-43b is consistent with a much older planet. This discrepancy is also noted by \citet{husnoo2012} and may potentially be explained by tidal interactions between the large planet and low-mass star leading to an increased stellar rotation rate and therefore an artificially younger age based on age-period relations \citep{pont2009,poppenhaeger2014,brown2011}. The results from our own stellar activity analysis show that WASP-43 is consistent with the population of old ($\sim5$ Gyr) inactive stars.

Due to large distance and expected low activity, we obtained no FUV observations of WASP-43 with the G140L or G140M gratings. We obtained NUV spectra with G230L and G430L with S/N above the threshold for wavelengths of $\lambda>2600$ $\text{\AA}$, including a strong Mg II emission line.

We obtained 28 ks of \textit{XMM-Newton} observations (obs.ID 0871800101, PI France) of WASP-43. The target was detected on the EPIC pn detector and OM and we find no evidence of flaring activity.

The X-ray and FUV proxy for WASP-43 is the K7 dwarf HD 85512.

\subsection{L 678-39}
L 678-39 (GJ 357, TOI-562) is a M2.5 dwarf with a DR3 distance of 9 pc and a sub-Solar to Solar metallicity of $\text{[Fe/H]}=-0.12\pm0.16$ \citep{schweitzer2019}. Long stellar rotation period, low $\log{R'_{HK}}$ value of -5.37, and low X-ray flux place the star in a regime of old age and low activity \citep{lugue2019,modirrousta2020}. The L 678-39 system contains of 3 confirmed exoplanets consisting of one Earth-sized planet and two super-Earth planets \citep{lugue2019}. The Earth-sized planet L 678-39b (GJ 357b) has a mass of $M_p=1.84\pm0.31$ $M_\earth$ and radius $R_p=1.217^{+0.084}_{-0.083}$ $R_\earth$ and is the closest to the host star at a distance of $a_p=0.035\pm0.002$ AU and an orbital period of 3.93 d. This system, along with the other two M dwarf systems in our study, is of interest due to the ongoing debate regarding the habitability of Earth-like planets around M dwarf stars.

We obtained FUV observations with the COS G130M and G160M gratings as well as STIS G140M for the Lyman-$\alpha$ emission. The high spectral resolution of the COS gratings provided good S/N over the FUV emission regions and we find a strong Lyman-$\alpha$ emission and reconstruct the intrinsic profile using the MCMC method of section \ref{sec:lya}. The NUV spectrum was obtained with STIS G230L and G430L and breaches the S/N$>3$ threshold for wavelengths of $\lambda>2600$ $\text{\AA}$. We detect a significant Mg II emission in the G230L spectrum. 

We obtained a 33 ks archival \textit{XMM-Newton} observation (obs.ID 0840841501, PI Stelzer) in which L 678-39 was detected with the high sensitivity EPIC pn detector. We find no evidence of flaring activity.

The X-ray and FUV proxy for L 678-39 is the M1.5 dwarf GJ 832.

\subsection{L 98-59}
L 98-59 (TOI-175) is a M3 dwarf with a DR3 distance of 10 pc and a metallicity of $\text{[Fe/H]}=-0.5\pm0.5$ \citep{kostov2019,cloutier2019}. \citet{cloutier2019} report $\log{R'_{HK}}=-5.4\pm0.11$. Combined with a rotational period of $p_{rot}\approx78$ d, this indicates an old, low-activity star \citep{astudillo2017}.

The system consists of four confirmed Earth to sub-Neptune sized planets (L 98-59b,c,d \citet{kostov2019}; L 98-59d \citet{demangeon2021}) and has gained much interest in the few years since its original discovery, prompting several follow up studies and observation proposals \citep{l981,l982,l983,cloutier2019}. In this work we consider only L 98-59b. L 98-59b has a mass of $M_p=0.4^{+0.16}_{-0.15}$ $M_\earth$ \citep{demangeon2021}, radius of $R_p=0.80\pm0.05$ $M_\earth$, and period of 2.25 d \citep{kostov2019}.

We obtained observations of L 98-59 with the STIS G140L, G140M, G230L, and G430L gratings. We find FUV emission lines greater than the S/N$>3$ threshold in the G140L spectra and strong Lyman-$\alpha$ emission in the G140M observations. Thus, our FUV emission line fluxes are relatively well constrained and we were able to reconstruct the Lyman-$\alpha$ line using the MCMC method. Our measurements of the FUV emission line fluxes and X-ray luminosity are consistent with the previous findings of low chromospheric activity. The G230L NUV observations break the S/N threshold for wavelengths of $\lambda>2600$ $\text{\AA}$.

We obtained two new \textit{XMM-Newton} observations (obs.ID 0871800201 and 0871800301, PI France) of duration 23 ks and 29.1 ks, respectively. The target was detected in both the EPIC pn and MOS detectors as well as the OM. Despite the target's low activity level indicators mentioned above, we find flaring events in both EPIC and OM detections which occur $\sim24$ hr apart. These flares are discussed in detail in section \ref{sec:flare}.

The X-ray and FUV proxy star for L 98-59 is again GJ 832.

\subsection{LP 791-18}
LP 791-18 is an M6 dwarf with a DR3 distance of 26 pc and an approximately solar metallicity $\text{[Fe/H]}=-0.09\pm0.19$. Age estimates from $v\sin{i}$ provide a lower limit of $>5$ Gyr \citep{crossfield2019}. The star is host to two confirmed exoplanets: a super-Earth with $R_p=1.1$ $R_\earth$ and a sub-Neptune with $R_p=2.3$ $R_\earth$. The planets have assumed but unconfirmed masses of 2 $M_\earth$ and 7 $M_\earth$, respectively \citep{crossfield2019}. We consider only the innermost planet, LP 791-18b. At the time of its discovery, this system was the third coolest confirmed exoplanet-hosting star---second to Teegarden's Star and TRAPPIST-1---making it of great interest to study the dynamics of multi-planet systems around small, very cool stars \citep{crossfield2019}.

Due to the expected faintness of the target we obtained only NUV observations with the STIS G230L and G430L gratings. The G230L spectrum never breaks the S/N$>3$ threshold except in the Mg II region which shows a faint emission line. The G430L spectrum did not break the S/N threshold until wavelengths of $\lambda>3800$ $\text{\AA}$.

We obtained 23.79 ks of new \textit{Chandra} observation of LP 791-18 with the ACIS-S instrument (obs.ID 23320, PI France). Our analysis of the observation showed a non-detection.

The X-ray and FUV proxy for LP 791-18 is Proxima Centauri (GJ 551), a $\sim5$ Gyr M5.5 dwarf.

\section{UV Flux Measurements}
\label{sec:emission_fluxes}

\begin{deluxetable}{cCCC}[H]
\tablecaption{UV emission line flux measurements [ergs s$^{-1}$ cm$^{-2}$]\tablenotemark{a}\label{tab:fluxes}}
\tablehead{
\colhead{Star}  &   \colhead{F$_{\text{Si III}}\lambda1206$}  &   \colhead{F$_{\text{N V}}\lambda\lambda1239,1243$} &   \colhead{F$_{\text{C II}}\lambda1335$}}
\startdata
WASP-17 &   \text{---} &   <1.02\times10^{-17}   &   <3.64\times10^{-18}\\
HD 149026   &   \text{---} &   6.84\pm0.59\times10^{-16} &   3.64\pm0.13\times10^{-15}\\
WASP-127    &   <3.68\times10^{-15}   &   <2.66\times10^{-17}   &   <2.84\times10^{-17}\\
WASP-77A    &   5.30\pm1.27\times10^{-16} &   <4.95\times10^{-17}   &   1.93\pm0.07\times10^{-15}\\
TOI-193 &   <5.16\times10^{-15}   &   <3.55\times10^{-17}   &   6.061\pm0.45\times10^{-16}\\
HAT-P-26    &   <4.18\times10^{-15}   &   <2.60\times10^{-17}   &   <7.58\times10^{-17}\\
HAT-P-12    &   \text{---} &   \text{---} &   \text{---}\\
WASP-43 &   \text{---} &   \text{---} &   \text{---}\\
L 678-39    &   1.17\pm0.27\times10^{-16} &   1.29\pm0.30\times10^{-16} &   2.09\pm0.44\times10^{-16}\\
L 98-59 &   <2.40\times10^{-15}   &   6.478\pm0.40\times10^{-16} &   8.92\pm0.37\times10^{-16}\\
LP 791-18   &   \text{---} &   \text{---} &   \text{---}\\
\enddata
\tablenotetext{a}{Upper limit values represent non-detected emission lines. The upper limit value reported is the RMS of the continuum-subtracted line region. Entries with a solid horizontal line represent no available data.}
\end{deluxetable}

\begin{deluxetable}{cCCCC}[H]
\tablecaption{UV emission line flux measurements cont. [ergs s$^{-1}$ cm$^{-2}$]\tablenotemark{\scriptsize{a}}\label{tab:fluxescont}}
\tablehead{
\colhead{Star}  &   \colhead{F$_{\text{Si IV}}\lambda\lambda1394,1403$}  &   \colhead{F$_{\text{C IV}}\lambda\lambda1548,1551$}    &   \colhead{F$_{\text{Mg II}}\lambda\lambda2799,2803$}   &   \colhead{F$_{\text{Ly}\alpha}$\tablenotemark{\scriptsize{b}}}}
\startdata
WASP-17   &   <1.31\times10^{-17} &   <1.58\times10^{-16}   &   7.50\pm0.50\times10^{-15} &   2.85\times10^{-15}\\
HD 149026 &   1.92\pm0.11\times10^{-15} &   3.52\pm0.23\times10^{-15} &   1.65\pm0.08\times10^{-13} &   2.01\pm0.37\times10^{-14}\\
WASP-127 &   <3.20\times10^{-17}   &   <1.73\times10^{-17}   &   1.59\pm0.20\times10^{-14} &   4.96\times10^{-15}\\
WASP-77A &   2.36\pm0.08\times10^{-15} &   2.87\pm0.13\times10^{-15} &   2.20\pm0.10\times10^{-14} &   7.41\times10^{-15}\\
TOI-193 &   <3.35\times10^{-17} &   1.11\pm0.10\times10^{-15}   &   1.81\pm0.21\times10^{-14} &   5.09\times10^{-15}\\
HAT-P-26   &   <9.34\times10^{-18}   &   <3.99\times10^{-17}   &   <4.62\times10^{-16}   &   <2.87\times10^{-15}\\
HAT-P-12 &   \text{---} &   \text{---} &   5.14\pm0.09\times10^{-15} &   8.95\pm0.19\times10^{-15}\\
WASP-43 &   \text{---} &   \text{---} &   2.18\pm0.02\times10^{-14} &   3.31\times10^{-14}\\
L 678-39 &   5.30\pm7.41\times10^{-17} &   5.82\pm1.84\times10^{-16} &   2.29\pm0.03\times10^{-14} &   7.56\pm0.17\times10^{-14}\\
L 98-59 &   7.94\pm0.43\times10^{-16} &   3.12\pm0.10\times10^{-15} &   1.84\pm0.02\times10^{-14} &   5.57\pm0.11\times10^{-14}\\
LP 791-18 &   \text{---} &   \text{---} &   1.72\pm0.07\times10^{-15} &   7.59\times10^{-15}\\
\enddata
\tablenotetext{a}{Upper limit values represent non-detected emission lines. The upper limit value reported is the RMS of the continuum-subtracted line region. Entries with a solid horizontal line represent no available data.}
\tablenotetext{b}{Lyman-$\alpha$ was detected in HAT-P-26 but is reported as an upper limit because we were unable to reconstruct the line profile and relied on the upper limit of the Mg II flux value.}
\end{deluxetable}


\bibliography{MUSCLES_extension}
\bibliographystyle{aasjournal}



\end{document}